\documentclass[conference]{IEEEtran}
\usepackage{graphicx}
\usepackage{amsmath}
\usepackage{lipsum}
\usepackage{amsthm}
\usepackage{amsfonts}
\usepackage{caption}
\usepackage{blkarray}
\usepackage{mathtools}
\usepackage{siunitx}
\usepackage{amssymb}
\usepackage{subcaption}
\usepackage{placeins}
\usepackage{lettrine}
\usepackage{multicol, blindtext}
\usepackage{bbm}
\usepackage{cite}
\usepackage{algorithm}
\usepackage{hyperref}
\usepackage{algpseudocode}
\usepackage{authblk}
\newtheorem{remark}{Remark}
\makeatletter
\newcommand\blfootnote[1]{%
  \begingroup
  \renewcommand\thefootnote{}\footnote{#1}%
  \addtocounter{footnote}{-1}%
  \endgroup
}

\newcommand{\Rmnum}[1]{\expandafter\@slowromancap\romannumeral #1@}
  \newenvironment{changemargin}[2]{\begin{list}{}{%
\setlength{\topsep}{0pt}%
\setlength{\leftmargin}{0pt}%
\setlength{\rightmargin}{0pt}%
\setlength{\listparindent}{\parindent}%
\setlength{\itemindent}{\parindent}%
\setlength{\parsep}{0pt plus 1pt}%
\addtolength{\leftmargin}{#1}%
\addtolength{\rightmargin}{#2}%
}\item }{\end{list}}
\newtheorem{theorem}{Theorem}
\newtheorem{corollary}{Corollary}
\newtheorem{proposition}{Proposition}
\newtheorem{assumption}{Assumption}
\newtheorem{definition}{Definition}
\newtheorem{lemma}{Lemma}
\newcommand{\bigzero}{\mbox{\normalfont\Large\bfseries 0}}
\newcommand{\bigb}{\mbox{\normalfont\Large\bfseries B}}

\DeclareMathOperator*{\argmax}{arg\,max}

\makeatletter
\def\BState{\State\hskip-\ALG@thistlm}
\makeatother
\ifCLASSINFOpdf
\else
\fi

\begin{document}
%
\title{On The Optimality of The Whittle's Index Policy For Minimizing The Age of Information}
\author[*]{Ali Maatouk}
\author[*]{Saad Kriouile}
\author[*]{Mohamad Assaad}
\author[$\dagger$]{Anthony Ephremides}
\affil[*]{TCL Chair on 5G, Laboratoire des Signaux et Syst\`emes, CentraleSup\'elec, Gif-sur-Yvette, France }
\affil[$\dagger$]{ECE Dept., University of Maryland, College Park, MD 20742}
\maketitle
\begin{abstract}
In this paper, we consider the average age minimization problem where a central entity schedules $M$ users among the $N$ available users for transmission over unreliable channels. It is well-known that obtaining the optimal policy, in this case, is out of reach. Accordingly, the Whittle's index policy has been suggested in earlier works as a heuristic for this problem. However, the analysis of its performance remained elusive. In the sequel, we overcome these difficulties and provide rigorous results on its asymptotic optimality in the
many-users regime. Specifically, we first establish its optimality in the neighborhood of a specific system's state. Next, we extend our proof to the global case under a recurrence assumption, which we verify numerically. These findings showcase that the Whittle's index policy has analytically provable optimality in the many-users regime for the AoI minimization problem. Finally, numerical results that showcase its performance and corroborate our theoretical findings are presented.
\blfootnote{This work has been supported by the TCL chair on 5G, ONR N000141812046, NSF CCF1813078, NSF CNS1551040, and NSF CCF1420651.}
\blfootnote{The first two authors contributed equally to this work.}
\end{abstract}
\IEEEpeerreviewmaketitle
\section{Introduction}
In the last decade, technological advances in the areas of sensors and wireless communications led to the emergence of a variety of new applications. Among the diverse functionalities provided by these applications, we cite monitoring, tracking, and even controlling operations in the physical world. For example, sensors can be interconnected to report environmental conditions or provide crucial data, such as the velocity and position in vehicular networks. To achieve the best performance, these applications rely heavily on the timely delivery of the data involved. To quantify this notion of timeliness, the Age of Information (\textbf{AoI}) was proposed \cite{6195689}. The AoI measures the information time-lag at the monitor side and, accordingly, its minimization is regarded as a means to achieve the freshness of information at the receiver side. 

Due to the large number of applications where timeliness of information is required, the AoI has captured a lot of research attention in recent years \cite{6195689,8000687,8262777,8445981,8006593,2019arXiv190102873Z,2019arXiv191003564B,8613408,8764468}. Since its inception, the AoI has been investigated in typical First-Come First-Served queuing settings \cite{6195689}. Packet management and the discard of stale packets were shown to further reduce the AoI in \cite{6875100}. Additionally, a wide range of applications were explored using the AoI as a performance metric \cite{8437927,8006542}. In the framework of remote estimation, new performance measures influenced by the AoI were proposed \cite{8406891,2019arXiv190706604M}. For example, the age of incorrect information was proposed in \cite{2019arXiv190706604M}. Priority-based queuing has also been extensively studied in the
age literature and is still gaining recent research attention \cite{8437591,8849695,2019arXiv190612278X}. 

Scheduling problems, which aim to minimize the AoI, have been widely explored in the literature \cite{Kadota:2018:SPM:3311528.3311547,2019arXiv190103020M,2019arXiv190107069Z,2018arXiv180103975J,2019arXiv190100481M}. For example, distributed scheduling solutions were proposed in \cite{2018arXiv180103975J,2019arXiv190100481M}. Specifically, back-off timers were optimized to minimize the average age in CSMA environments in \cite{2019arXiv190100481M}. Among the scheduling problems tackled in the literature, we cite the following: consider $N$ users communicating with a central entity over unreliable channels where at most $M$ users can transmit simultaneously. What is the age-optimal strategy in this case? This problem is considered a fundamental one due to its wide
range of applications and, therefore, 
has been
previously investigated in \cite{Kadota:2018:SPM:3311528.3311547}. The authors in \cite{Kadota:2018:SPM:3311528.3311547} have shown that a greedy algorithm is optimal when users have identical channel statistics. In the asymmetric case, sub-optimal policies were proposed, one of which is the Whittle's index policy \cite{Kadota:2018:SPM:3311528.3311547}. 
%
%
%

The Whittle's index policy revolves around: 1) assigning an index to each user based on its age and channel statistics, 2) scheduling the $M$ users with the highest indices. This policy has been previously adopted for a large variety of scenarios (e.g., delay minimization \cite{8849774,2018arXiv180700352K} and throughput maximization \cite{ouyang2016downlink,8115326,7606842}). The main interest in this policy stems from its simplicity and notable performance\cite{2018arXiv180700352K,8115326,7606842,8849774,ouyang2016downlink,weber1990index}. However, the analysis of its performance is known to be challenging and, accordingly, remains elusive in the average age minimization framework. In fact, thus far in the age literature, a bound on its performance has been developed in \cite{Kadota:2018:SPM:3311528.3311547}, and numerical implementations of this policy were done for several scenarios (e.g., \cite{2019arXiv190810438T,2017arXiv171207419H,8935400}). In the sequel, we overcome these difficulties and provide rigorous analytical results on its performance. More specifically, we prove that the aforementioned policy is age-optimal for the general asymmetrical case in the
many-users regime. The many-users settings present themselves as an interesting regime to explore due to the astronomical growth in the number of interconnected devices. It is forecasted that over 25 billion IoT devices will be installed by 2020, mainly through wireless networks. For example, machine-type communications and the IoT in 5G networks require supporting tens of thousands of connected devices in a single cell. Therefore, the characterization of the optimal scheduling policies in this asymptotic regime is of paramount importance. Knowing that the freshness of information is of broad interest in IoT applications, we emphasize on the importance of the analytical results laid out in our paper. To that end, the following are our key contributions:
\begin{itemize}
\item First, we propose a modified version of the standard age metric. This version will provide us with analytical benefits throughout our paper. We note that this proposed measure can be made arbitrarily close to the traditional age metric as desired.
\item Afterward, we formulate the problem of minimizing the average modified age of the network when $M$ among $N$ users can communicate with a central entity simultaneously. Since the problem belongs to the family of Restless Multi-Armed Bandit (\textbf{RMAB}) problems, finding the optimal policy is known to be out of reach. Accordingly, we present a relaxed version of the problem and tackle it through a Lagrangian approach. Subsequently, we find the Whittle's index expressions and establish the Whittle's index policy for the original scheduling problem.
\item Next, we introduce a fluid limit model to approximate the behavior of the system when the Whittle's index policy is employed. As $N$ increases, we prove that the fluid limit can be made arbitrarily close to the actual system state evolution. Therefore, we mainly focus on the evolution of the fluid limit vector in our analysis. By leveraging matrices tools, we show that this vector converges to a unique fixed point. We use these results to establish the optimality of the Whittle's index policy in the neighborhood of the fixed point. Finally, under a recurrence assumption which we verify numerically, we extend our results to any random initial system state. These results put into perspective the analytically proven optimality of the Whittle's index policy. Lastly, we provide numerical results that corroborate these theoretical findings.
\end{itemize}
The rest of the paper is organized as follows: Section II is
dedicated to the system model. Section III incorporates our analysis of the scheduling problem where we provide key structural results on its optimal solution and establish the Whittle's index scheduling policy. In Section IV and V, we present the main contribution of the paper where we prove the asymptotic optimality of the Whittle's index policy. Numerical results that corroborate
these findings are given in Section VI while Section VII concludes the paper.
%
%
\section{System Model}
\subsection{Network description}
We consider in our paper $N$ users that generate and send status updates to a central entity. The goal is to let the central entity have the freshest knowledge possible of the information carried by each user. Due to the limited amount of resources, only $M<N$ users can transmit simultaneously and a scheduling scheme has to be adopted to achieve the aforementioned goal. Let $\alpha=\frac{M}{N}\in]0,1[$ be the portion of the $N$ users that can transmit simultaneously. Time is considered to be discrete and normalized to the time slot duration (i.e., $t=0,1,2,\ldots$). At time $t$, if user $i$ is among the scheduled users, this user generates a packet and sends it to the central entity during this time slot. The transmitted packet is successfully decoded by the receiver at time $t+1$ with a probability $p_i$ and a transmission error occurs with probability $1-p_i$. In practice, users may share similar channel conditions. Accordingly, we suppose that users are divided into $K$ classes such that the probability of successful transmission for users in class $k$ is $p_k$. Moreover, each class $k$ has $\gamma_kN$ users and, consequently, the following holds: $\sum_{k=1}^{K}\gamma_k=1$. 
\subsection{Adopted age metric}
At each time instant $t\geq 0$, we let $B_{i}^k(t)$ be the timestamp of the freshest packet by user $i$ of class $k$ that has been delivered to the central entity. The age of information, or simply the age, of the aforementioned user is defined as \cite{6195689}:
\begin{equation}
\Delta_{i}^k(t)=t-B_{i}^k(t)
\label{agedef}
\end{equation}
We adopt in the sequel a modified version of the age in (\ref{agedef}). Specifically, we let the modified age of user $i$ of class $k$ be: 
\begin{equation}
S_{i}^{k}(t)=\min(\Delta_{i}^k(t),L)
\label{newdef}
\end{equation}
where $L$ is a \emph{finite} positive integer in $\mathbb{N}^*$. It can be easily seen from the definitions that the two metrics coincide for large values of $L$. This modified definition of the age will be of great analytical importance in the remainder of the paper. 
\subsection{Problem formulation}
We let the age vector at time $t$ be $\boldsymbol{S}(t)=(S_{1}^1(t),\ldots,S_{\gamma_KN}^{K}(t))$ where $S_i^k(t)$ is the \emph{modified} age at the central entity of user $i$ of class $k$ at time slot $t$. The objective is to find a scheduling policy that minimizes the expected total average age of the network. A scheduling policy $\pi$ is defined as a sequence of actions $\pi=(\boldsymbol{a}^{\pi}(0),\boldsymbol{a}^{\pi}(1),\ldots)$ where $\boldsymbol{a}^{\pi}(t)=(a_1^{1,\pi}(t),a_2^{1,\pi}(t),\ldots,a_{\gamma_KN}^{K,\pi}(t))$ is a binary vector such that $a_i^{k,\pi}(t)=1$ if user $i$ of class $k$ is scheduled at time $t$. We let $h_i^{k,\pi}(t)\in\{0,1\}$ be an i.i.d. Bernoulli random variable that indicates if the transmitted packet by the scheduled user is successfully received (value $1$) or a transmission error took place (value $0$). As it was previously explained, we have $\Pr(h_i^{k,\pi}(t)=1)=p_k$ and $\Pr(h_i^{k,\pi}(t)=0)=1-p_k$. Accordingly, the evolution of the age of user $i$ of class $k$ under policy $\pi$ can be summarized in the following: 
\begin{equation*}
S_i^k(t+1) = \begin{cases}
              1 & \text{if} \:\:  a_i^{k,\pi}(t)=1,h_i^{k,\pi}(t)=1 \\
             
             \min(S_i^k(t)+1,L)  & \text{otherwise}
       \end{cases} 
\label{funct}
\end{equation*}
Note that the age state space is finite since $S_i^k(t)\in\{1,\ldots,L\}$ for any $t\geq0$. By letting $\Pi$ be the set of all \emph{causal} scheduling policies,  
our scheduling problem can formulated as follows:
\begin{equation}
\setlength{\belowdisplayskip}{0pt} \setlength{\belowdisplayshortskip}{0pt}
\setlength{\abovedisplayskip}{0pt} \setlength{\abovedisplayshortskip}{0pt} 
\begin{aligned}
& \underset{\pi\in \Pi}{\text{minimize}}
& & \lim_{T\to+\infty} \text{sup}\:\frac{1}{T}\mathbb{E}^{\pi}\Big(\sum_{t=0}^{T-1}\sum_{k=1}^{K}\sum_{i=1}^{\gamma_kN}S_i^{k,\pi}(t)|\boldsymbol{S}(0)\Big)\\
& \text{subject to}
& & \sum_{k=1}^{K}\sum_{i=1}^{\gamma_kN}a_{i}^{k,\pi}(t)\leq\alpha N \quad t=1,2,\ldots
\end{aligned}
\label{originalobjective}
\end{equation}
The problem in (\ref{originalobjective}) belongs to the family of RMAB problems. In multi-armed bandit problems, at each decision epoch, a
scheduler chooses which bandit
to play, and a penalty is incurred accordingly. In the restless case, the states of all bandits (the age of users in our case) evolve even when these bandits are not chosen. The objective is to design a bandit selection policy that minimizes the average expected penalty. However, obtaining this optimal policy is known to be out of reach. A well-known heuristic for this type of problems is the Whittle's index policy \cite{weber1990index}. This policy is based on a Lagrangian relaxation, and was shown to have remarkable performance in real-life applications. 
\section{Lagrangian Relaxation and Whittle's Index}
\subsection{Relaxed problem}
The Lagrangian relaxation technique presents itself as a main component for defining the Whittle's index scheduling policy. First, it consists of relaxing the constraint on the available resources by letting it be satisfied on average rather than in every time slot. More specifically, we define our Relaxed Problem (\textbf{RP}) as follows:
\begin{equation}
\setlength{\belowdisplayskip}{0pt} \setlength{\belowdisplayshortskip}{0pt}
\setlength{\abovedisplayskip}{0pt} \setlength{\abovedisplayshortskip}{0pt} 
\begin{aligned}
& \underset{\pi\in \Pi}{\text{minimize}}
& & \lim_{T\to+\infty} \text{sup}\:\frac{1}{T}\mathbb{E}^{\pi}\Big(\sum_{t=0}^{T-1}\sum_{k=1}^{K}\sum_{i=1}^{\gamma_kN}S_i^{k,\pi}(t)|\boldsymbol{S}(0)\Big)\\
& \text{subject to}
& & \lim_{T\to+\infty}\frac{1}{T}\mathbb{E}^{\pi}\Big(\sum_{t=0}^{T-1}\sum_{k=1}^{K}\sum_{i=1}^{\gamma_kN}a_{i}^{k,\pi}(t)\Big)\leq\alpha N
\end{aligned}
\label{relaxedobjective}
\end{equation}
Afterward, the constraint is incorporated in the objective function using the Lagrangian function $f(W,\pi)$:
\begin{equation}
\lim_{T\to+\infty} \text{sup}\:\frac{1}{T}\mathbb{E}^{\pi}\Big(\sum_{t=0}^{T-1}\sum_{k=1}^{K}\sum_{i=1}^{\gamma_kN}S_i^{k,\pi}(t)+Wa_{i}^{k,\pi}(t)|\boldsymbol{S}(0)\Big)-W\alpha N
\end{equation}
where $W\geq0$ can be seen as a penalty for scheduling users.  Accordingly, by following the Lagrangian approach, our next goal is to solve the following problem:
\begin{equation}
\setlength{\belowdisplayskip}{0pt} \setlength{\belowdisplayshortskip}{0pt}
\setlength{\abovedisplayskip}{0pt} \setlength{\abovedisplayshortskip}{0pt} 
\begin{aligned}
& \underset{\pi\in \Pi}{\text{minimize}}
& & f(W,\pi)
\end{aligned}
\label{relaxedobjectivelagrange}
\end{equation}
As the term $W\alpha N$ is independent of $\pi$, it can be omitted from the analysis. With the above formulation in mind, we present the general recipe to obtain the Whittle's index policy:
\begin{enumerate}
\item We focus on the one-dimensional version of the problem in (\ref{relaxedobjectivelagrange}). It can be shown that the $N$-dimensional problem can be decomposed into $N$ one-dimensional problems that can be solved independently \cite{2018arXiv180700352K}. Consequently, we drop the user's index (and any class-dependent indices) for ease of notation, and we focus on \emph{one instance} of these one-dimensional problems:
\begin{equation}
\setlength{\belowdisplayskip}{0pt} \setlength{\belowdisplayshortskip}{0pt}
\setlength{\abovedisplayskip}{0pt} \setlength{\abovedisplayshortskip}{0pt} 
\begin{aligned}
& \underset{\pi\in \Pi}{\text{minimize}}
& & \lim_{T\to+\infty} \text{sup}\:\frac{1}{T}\mathbb{E}^{\pi}\Big(\sum_{t=0}^{T-1}S^{\pi}(t)+Wa^{\pi}(t)|\boldsymbol{S}(0)\Big)
\end{aligned}
\label{onedimensional}
\end{equation}
\item We provide structural results on the optimal solution of the one-dimensional problem.
\item We establish the indexability property, which
ensures the existence of the Whittle's indices.
\item We derive a closed-form expression of the Whittle's index and, accordingly, define the proposed scheduling policy for the original problem (\ref{originalobjective}).
\end{enumerate}
\subsection{Structural results}
Based on our model's assumptions and the dynamics previously detailed in Section II-C, the problem in (\ref{onedimensional}) can be cast into an infinite horizon average cost Markov decision process that is defined as follows:
\begin{itemize}
\item \textbf{States}: The state of the MDP at time $t$ is nothing but the penalty function $S(t)$. The penalty can have any value in $\{1,\ldots,L\}$. Therefore, the considered state space is finite.
\item \textbf{Actions}: The action at time $t$, denoted by $a(t)$, indicates if the user is scheduled (value $1$) or not (value $0$).
\item \textbf{Transitions probabilities}: The transitions probabilities between the different states have been previously detailed in Section II-C.
\item \textbf{Cost}: We let the instantaneous cost of the MDP, $C(S(t),a(t))$, be equal to $S(t)+Wa(t)$.
\end{itemize}
Finding the optimal solution of an infinite horizon average cost MDP is known to be challenging due to the curse
of dimensionality. Specifically, it is well-known that the optimal policy $\pi^*$ of the one-dimensional problem can be obtained by solving the following Bellman equation for all states $S$:
\begin{equation}
\theta + V(S)=\min_{a\in\{0,1\}}\big\{S+Wa+\sum_{S'\in\{1,\ldots,L\} }\Pr(S\rightarrow S'|a)V(S')\big\} \quad 
\label{bellman}
\end{equation}
where $\Pr(S\rightarrow S')$ is the transition probability from state $S$ to $S'$, $\theta$ is the optimal value of the problem and $V(S)$ is the value function. Based on (\ref{bellman}), one can see that the optimal policy $\pi^*$ depends on $V(.)$, for which
there is no closed-form solution in general. There exist various numerical algorithms that solve (\ref{bellman}), such as the value and policy iteration algorithms. However, they suffer from being computationally demanding. To circumvent this complexity, we focus on studying the structure of the optimal scheduling policy. By doing so, the following results can be obtained. 
\begin{definition}
An increasing threshold policy is a deterministic stationary policy in which we do not schedule a user if its current state $S$ is smaller than a certain threshold $n\in\{1,\ldots, L\}$. On the other hand, if $S$ is greater or equal than $n$, the user is scheduled.
\end{definition}
\begin{theorem}
The optimal solution of the problem in (\ref{onedimensional}) is an increasing threshold policy.
\label{increasingthreshold}
\end{theorem}
\begin{IEEEproof}
The proof can be found in Appendix \ref{proofoftheoremincreasingthreshold}.
\end{IEEEproof}
\subsection{Indexability and Whittle's index expressions}
\begin{figure}[!ht]
\centering
\includegraphics[width=.99\linewidth]{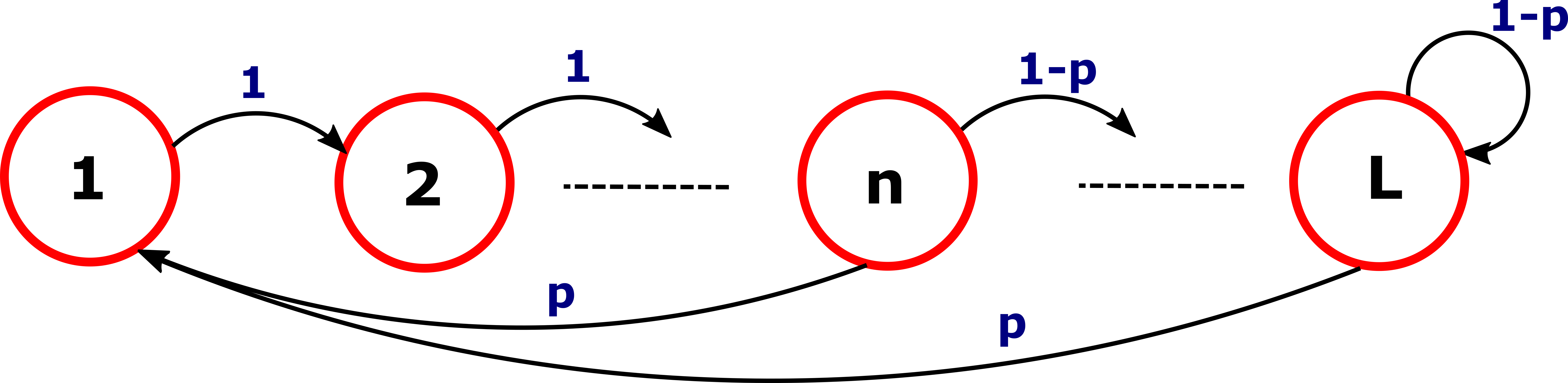}
\setlength{\belowcaptionskip}{-5pt}
\caption{The states transitions under a threshold policy}
\label{thresholddtmc}
\end{figure}

In order to establish the indexability of the problem and find the Whittle's index expressions, we tackle in depth the behavior of the MDP when a threshold policy is adopted. To that end, we recall that a threshold policy is fully characterized by its threshold value $n$. Consequently, the problem in (\ref{onedimensional}) can be reformulated as follows:
\begin{equation}
\begin{aligned}
& \underset{n\in \mathbb{N}^*}{\text{minimize}} 
& & \overline{C}(n,W)
\end{aligned}
\label{thresholdobjective}
\end{equation}
where $\overline{C}(n,W)$ is the average cost of the MDP when the threshold policy is adopted. We note that for any threshold $n\in\{1,\ldots,L\}$, the MDP can be modeled through a Discrete Time Markov Chain (\textbf{DTMC}) as seen in Fig. \ref{thresholddtmc} where:
\begin{itemize}
\item The states refer to the values of the penalty function $S(t)$.
\item For any state $S(t)<n$, the user is not scheduled. On the other hand, for any state $S(t)\geq n$, the user is scheduled and the dynamics of $S(t)$ coincide with those previously reported in Section II-C. 
\end{itemize}
The next step in our analysis consists of calculating the average cost function $\overline{C}(n,W)$. To do so, we find the stationary distribution of the DTMC in question. 
\begin{proposition}
For any given threshold $n\in\{1,\ldots,L\}$, the DTMC is irreducible and admits $u^n(i)$ for $i=1,\ldots,L$ as its stationary distribution where\footnote{For any threshold $n\geq L+1$, $u^n(i)=0$ for any $i<L$ and $u^n(L)=1$.}:
\begin{equation}
 u^n(i)=\left\{
    \begin{array}{ll}
        \frac{p}{np+1-p} & \text{if} \ 1 \leq i \leq n  \\
        (1-p)^{i-n} \frac{p}{np+1-p} & \text{if} \ i \geq n\\
        \frac{(1-p)^{L-n}}{np+1-p} & \text{if} \ i=L
    \end{array}
\right.
\end{equation}
\label{stationarydistribution}
\end{proposition}
\vspace{-10pt}
\begin{IEEEproof}
The proof can be found in Appendix \ref{proofofstationary}.
\end{IEEEproof}
Note that, unless otherwise specified, we will use the index $i=1,\ldots,L$ in the sequel to denote the age state. By leveraging the above results, we can proceed with finding a closed-form of the average cost of any threshold policy.
\begin{theorem}
For any given threshold $n\in\{1,\ldots,L\}$, the average cost of the threshold policy is $\overline{C}(n,W)=\overline{C}_1(n)+\overline{C}_2(n,W)$ where\footnote{For any threshold $n\geq L+1$, $\overline{C}_1(n)=L$ and $\overline{C}_2(n,W)=0$.}:
\begin{align}
\overline{C}_1(n)=&\frac{[(n-1)^2+(n-1)]p^2+2p(n-1)}{2p((n-1)p+1)}\nonumber\\&+\frac{2[1-(1-p)^{L-n+1}]}{2p((n-1)p+1)}
\label{costgamma}
\end{align}
\begin{equation}
\overline{C}_2(n,W)=\frac{W}{np+1-p}
\label{costlambda}
\end{equation}
\label{theoremcostgamma}
\end{theorem}
\vspace{-15pt}
\begin{IEEEproof}
The proof can be found in Appendix \ref{proofofcostgamma}.
\end{IEEEproof}
Next, we establish the indexability property of the problem for all users, which ensures the existence of the Whittle's indices and allows us to establish our index policy.
\begin{definition}[Indexability]
For a fixed $W$, consider the vector $\boldsymbol{l}(W)=(l_1(W),\ldots,l_K(W))$ where $l_k(W)$ is the optimal threshold for the problem in (\ref{thresholdobjective}) for each user of class $k$. We define $D^k(W)=\{S\in\{1,\ldots,L\}:S<l_k(W)\}$ as the set of states for which the optimal action is to not schedule the users belonging to class $k$. The one-dimensional problem associated with these users is said to be indexable if $D^k(W)$ is increasing in $W$. More specifically, the following should hold:
\begin{equation}
W'\leq W \Rightarrow D^k(W') \subseteq D^k(W)
\end{equation}
\label{indexabilitydefinition}
\end{definition}
\vspace{-25pt}
\begin{proposition}
For each user belonging to any class $k=1,\ldots,K$, the one-dimensional problem is indexable.
\label{indexabilityproperty}
\end{proposition}
\begin{IEEEproof}
The proof can be found in Appendix \ref{proofofindexability}.
\end{IEEEproof}
As the indexability property has been established in the above proposition, we can now affirm the existence of the Whittle's index. To that end, we first define the Whittle's index and then provide a closed-form expression of it.
\begin{definition}
The Whittle's index $W_i^k$ of a state $i$ of class $k$ is defined as the infimum subsidy $W$ that makes both idling and transmitting equally desirable in state $i$ of class $k$.
\end{definition}
\begin{proposition}
For any class $k$ and state $i=1,\ldots,L$, the Whittle's index is:
\begin{equation}
W^k_i=\frac{i(i-1)p_k}{2}+i-i(1-p_k)^{L-i}
\label{whittleindicesexpression}
\end{equation}
\label{closedformwhittle}
\end{proposition}
\vspace{-15pt}
\begin{IEEEproof}
The proof can be found in Appendix \ref{proofofclosedform}.
\end{IEEEproof}
With the Whittle's index expression being found, we summarize in the following the Whittle's index scheduling policy for the original problem (\ref{originalobjective}).
\begin{algorithm}
\caption{Whittle's index scheduling policy}\label{euclid}
\begin{algorithmic}[1]
\State At each time slot $t$, calculate the Whittle's index of all users in the network using (\ref{whittleindicesexpression}).
\State Schedule the $M$ users having the highest Whittle's index values at time $t$, with ties broken arbitrarily.
\end{algorithmic}
\end{algorithm}\\
Although the above scheduling policy is easy to implement, it remains sub-optimal. Accordingly, characterizing its performance compared to the optimal policy is important.
\section{Local optimality}
In this section, we study the local optimality properties of the
Whittle's index policy. The results of this section will be the basis of our subsequent analysis of global optimality. To that end, we start by introducing a state space over which the local optimality will be established.
\subsection{System state}
We denote by $Z_i^{k,N}(t)$ the proportion of users of class $k$ in state $i$ at time $t$ when $N$ users are in the network. For example, if at time $t$, all users of class $k$ have an age equal to $1$ then $Z_1^{k,N}(t)=\gamma_k$ and $Z_i^{k,N}(t)=0$ for any $i\neq1$. By considering all the classes in the network, we define the system state vector $\boldsymbol{Z}^N(t)$:
\begin{equation}
\boldsymbol{Z}^N(t)=(\boldsymbol{Z}^{1,N}(t),\ldots,\boldsymbol{Z}^{K,N}(t))
\end{equation}
where $\boldsymbol{Z}^{k,N}(t)=(Z_1^{k,N}(t),\ldots,Z_L^{k,N}(t))$ is the state vector for class $k$. We define the state space to which $\boldsymbol{Z}^N(t)$ belongs as follows:
\begin{equation}
\mathcal{Z}=\{\boldsymbol{Z}^N\geq0:\sum\limits_{i=1}^{L}Z_i^{k,N}(t)=\gamma_k \quad k=1,\ldots,K\}
\end{equation}
For any scheduling policy $\pi$, the system state vector $\boldsymbol{Z}^N(t)$ evolves depending on the actions taken by the scheduler. By characterizing the evolution of the system state vector $\boldsymbol{Z}^N(t)$ when the Whittle's index policy is employed, we will be able to establish its optimality. To proceed in that direction, we first specify the optimal solution of the relaxed problem in (\ref{relaxedobjective}).
\subsection{Optimal solution of the Relaxed Problem}
In the previous section, we have established that problem (\ref{relaxedobjectivelagrange}) can be solved optimally for any $W$ using a threshold policy and we have noted by $\boldsymbol{l}(W)=(l_1(W),\ldots,l_K(W))$ this optimal threshold vector. With that in mind, the question that should be answered is the following: what is the optimal policy of the relaxed problem (\ref{relaxedobjective})? To answer it, we first seek to find $\boldsymbol{l}(W)$ in function of the Whittle's index.
\begin{proposition}[Optimal Threshold]
For any given $W$, each element $l_k(W)$ of the optimal threshold vector $\boldsymbol{l}(W)$ has one
of the following two expressions\footnote{If $\{W_{i}^k:W_{i}^k\leq W\}$ or $\{W_{i}^k:W_{i}^k< W\}$ is empty, we consider that $\underset{i=1,\ldots,L}{\argmax} \{W_{i}^k:W_{i}^k\leq W\}$, resp. $\underset{i=1,\ldots,L}{\argmax} \{W_{i}^k:W_{i}^k<W\}$, is equal to $0$.}:
\begin{equation*}
l_k^1(W)=\underset{i=1,\ldots,L}{\argmax} \{W_{i}^k:W_{i}^k\leq W\}+1 \quad\forall k\in\{1,\ldots,K\}
\end{equation*}
\begin{equation}
l_k^2(W)=\underset{i=1,\ldots,L}{\argmax} \{W_{i}^k:W_{i}^k<W\}+1 \quad\forall k\in\{1,\ldots,K\}
\label{twoexpressions}
\end{equation}
\label{lwvector}
where $W_{i}^k$ is the Whittle's index of a state $i$ in class $k$.
\end{proposition}
\begin{IEEEproof}
The proof can be found in Appendix \ref{proofoflwvector}.
\end{IEEEproof}
With the above results in mind, we now identify the optimal solution of the relaxed problem (\ref{relaxedobjective})
%
\begin{proposition}[Optimal Policy of the RP]
The optimal policy of the relaxed problem (\ref{relaxedobjective}) is a threshold policy, characterized by a real value $W^*\in\mathbb{R}^+$ and a randomization parameter $\theta^*\in[0,1]$ such that:
\begin{itemize}
\item There exists a class $m$ and state $p$ such that $W^*=W_p^m$.
\item The threshold vector $\boldsymbol{l}(W^*)$ coincides with the expressions provided in Proposition \ref{lwvector}.
\item The thresholds $l^2_m(W^*)$ and $l^1_m(W^*)$ are used for users belonging to class $m$ with a probability $\theta^*$ and $1-\theta^*$ respectively.
\item $W^*$ and $\theta^*$ are chosen in a way that the expected proportion of scheduled users is exactly equal to $\alpha$.
\end{itemize}
%
%
%
%
%
\label{optimalrelaxed}
\end{proposition} 
\begin{IEEEproof}
The proof can be found in Appendix \ref{proofoptimalrelaxed}.
\end{IEEEproof}
This characterization of the optimal policy of the RP will
allow us to find an expression of the optimal average age of
the relaxed problem, denoted by $C^{RP,N}$: 
\begin{align}
C^{RP,N}&=\sum_{\substack{k=1 \\ k\neq m}}^K \gamma_k N \sum_{i=1}^{L} u_k^{l_k(W^*)}(i)i+\gamma_m N\theta^*\sum_{i=1}^{L} u_m^{l_m^2(W^*)}(i)i\nonumber\\&+\gamma_m N(1-\theta^*)\sum_{i=1}^{L} u_m^{l_m^1(W^*)}(i)i
\label{expressionofoptimalcost}
\end{align}
where $u_k^{l_k(W^*)}(i)$ is the stationary distribution of state $i$ in class $k$ when the threshold $l_k(W^*)$ is employed. The importance of these results comes from the fact that
\begin{equation}
C^{RP,N}\leq C^{PP,N}\leq C^{WI,N}
\end{equation}
where $C^{PP,N}$ and $C^{WI,N}$ are the optimal average age of the primal problem (\ref{originalobjective}) and the average age when the Whittle's index policy is adopted respectively. Accordingly, if we prove that $C^{WI,N}$ converges to $C^{RP,N}$ when $N$ grows, then it is surely optimal for the original problem (\ref{originalobjective}).
\subsection{Fluid limit}
In this section, we present a fluid limit model to approximate the behavior of the system state vector $\boldsymbol{Z}^N(t)$ when the Whittle's index policy is employed. Specifically, we define a deterministic vector $\boldsymbol{z}(t)\in\mathcal{Z}$ that evolves as follows:
%
\begin{equation}
\boldsymbol{z}(t+1)-\boldsymbol{z}(t)|_{\boldsymbol{z}(t)=\boldsymbol{z}}=\mathbb{E}[\boldsymbol{Z}^N(t+1)-\boldsymbol{Z}^N(t)|\boldsymbol{Z}^N(t)=\boldsymbol{z}]
\label{expectationdefinition}
\end{equation}
where $\boldsymbol{z}\in\mathcal{Z}$ is any feasible system state. In a later part of the paper, we will characterize the gap between $\boldsymbol{Z}^N(t)$ and $\boldsymbol{z}(t)$, and we will show that this gap vanishes when the number of users is high. For now, we will focus on characterizing the evolution of $\boldsymbol{z}(t)$. To that end, let us consider that the system is in state $\boldsymbol{z}$ and we denote by $W_j^h$ the Whittle's index of users belonging to class $h$ having an age equal to $j$. We let $p_i^k(\boldsymbol{z})$ be the probability that a user of class $k$ in state $i$ is selected among the proportion $z_i^k$ for transmission. By following \cite{weber1990index}, one can show:
\begin{equation}
p_i^k(\boldsymbol{z})=\min \{ z_i^k,\max(0,\alpha-\sum_{W_j^h > W_i^k} z_j^h) \} /z_i^k
\label{pikz}
\end{equation}
where the summation is over the proportion of users of any class $h$ in state $j$ such that $W_j^h>W_i^k$. Next, given $\boldsymbol{z}$, we let $q_{i,j}^k(\boldsymbol{z})$ be the probability of transition between state $i$ and $j$ in class $k$. Moreover, we denote by $q_{i,j}^{k,0}$ and $q_{i,j}^{k,1}$ the probability of transition from state $i$ to state $j$ in a class $k$ if the user is left to idle or is scheduled for transmission respectively. Accordingly, we have:
\begin{equation}
q_{i,j}^k(\boldsymbol{z})=p_i^k(\boldsymbol{z})q_{i,j}^{k,1}+(1-p_i^k(\boldsymbol{z}))q_{i,j}^{k,0}
\label{qikz}
\end{equation}
Based on the above, and using the definition of the fluid limit, we can conclude that the evolution of $\boldsymbol{z}(t)$ can be summarized as follows:
\begin{equation}
z_i^k(t+1)-z_i^k(t)=\sum_{j \neq i} q_{j,i}^k(\boldsymbol{z}(t))z_j^k(t)-\sum_{i\neq j}q_{i,j}^k(\boldsymbol{z}(t))z_i^k(t)
\label{equationadime}
\end{equation}
This can be rewritten in the following manner:
\begin{equation}
\boldsymbol{z}(t+1)=Q'(\boldsymbol{z}(t))\boldsymbol{z}(t)
\label{zelases}
\end{equation}
where $Q'(\boldsymbol{z}(t))=Q(\boldsymbol{z}(t))+I$ and $I$ is the identity matrix of dimension $LK$. Investigating (\ref{zelases}) for any state $\boldsymbol{z}\in\mathcal{Z}$ is challenging since $Q'$ depends on $\boldsymbol{z}$. To overcome this, we restrict our analysis to a specific set in $\mathcal{Z}$ as we will detail in the following. Let us consider $W^*$ and $\boldsymbol{l}(W^*)$ previously detailed in Proposition \ref{optimalrelaxed}. 
In the sequel, and for ease of notation, we do the following:
\begin{itemize}
\item For $k\neq m$, we have shown in Appendix \ref{proofoflwvector} that $l_k(W^*)=l_k^1(W^*)=l_k^2(W^*)$. Accordingly, we will refer to this threshold simply by $l_k^*$.
\item We will refer to $l^2_m(W^*)$ by $l^*_m$.
\end{itemize}
We define $\jmath_{W^*}$ as the set of vectors $\boldsymbol{z}\in\mathcal{Z}$ such that users with a Whittle's index strictly larger than $W^*$ are scheduled. On the other hand, users with a Whittle's index strictly smaller than $W^*$ are left to idle. Meanwhile, users that have a Whittle's index equal to $W^*$ are scheduled with a certain randomization. Accordingly, we can write $\jmath_{W^*}$ as follows:
\begin{equation}
\jmath_{w^*}=\{\boldsymbol{z}\in\mathcal{Z}: \ \sum_{W_i^k>W^*} z_i^k<\alpha, \sum_{W_i^k \geq W^*} z_i^k \geq \alpha \}
\end{equation}
We recall that for any class $k$ and time $t\geq0$, $\sum_{j=1}^L z_j^k(t)=\gamma_k$. Accordingly, for $k\neq m$, we can replace $z_{l_k^*-1}^k(t)$ by $\gamma_k-\sum_{j=1, j\neq {l_k^*-1}}^L z_j^k(t)$. Therefore, there is no need to track the evolution of $z_{l_k^*-1}^k(t)$ with time as it can deduced from the evolution of $z_j^k(t)$ for $j\neq l_k^*-1$. Accordingly, we let $\widetilde{\boldsymbol{z}}^k=[z_1^k,\ldots,z_{l_k^*-2}^k(t),z_{l_k^*}^k(t),\ldots,z_{L}^k(t)]$. Next, by replacing $q_{j,i}^{k}(z(t))$ with its value in (\ref{equationadime}), we can obtain the following relationship between $\boldsymbol{z}(t+1)$ and $\boldsymbol{z}(t)$ for any $\boldsymbol{z}(t) \in \jmath_{w^*}$.\\
1) $k\neq m$: In this case, $z_i^k(t+1)$ will have the following expression:
\begin{equation}
    \begin{array}{ll}
        \sum_{j=l_k^*}^L p_k z_j^k(t) & \text{if} \ i=1  \\
        z_{i-1}^k(t) & \text{if} \ 1< i < l_k^*-1 \\
        -\sum_{j=1}^{l_k^*-2} z_j^k(t)-\sum_{j=l_k^*}^L z_j^k(t)+\gamma_k & \text{if} \ i=l_k^*\\
        (1-p_k)z_{i-1}^k(t) & \text{if} \ l_k^*< i <L\\ 
        (1-p_k) z_{L-1}^k(t)+(1-p_k)z_L^k(t) & \text{if} \ i=L
    \end{array}
\label{zktplusone}
\end{equation}
2) $k=m$: To tackle this case, we first replace $z_{l_m^*}^m(t)$ by $\gamma_m-\sum_{j=1, j\neq {l_m^*}}^L z_j^m(t)$. Therefore, there is no need to track the evolution of $z_{l_m^*}^m(t)$ with time. Accordingly, we let $\widetilde{\boldsymbol{z}}^m=[z_1^m,\ldots,z_{l_m^*-1}^m(t),z_{l_m^*+1}^m(t),\ldots,z_{L}^m(t)]$. Next, we recall that the portion of scheduled users in the set $\jmath_{w^*}$ is always equal to $\alpha$. Hence, the portion of scheduled users of class $m$ can be always written as the difference $\alpha-\sum_{k\neq m}\sum_{j=l_k^*}^L z_j^k(t)$. With that in mind, we can write the evolution of $z_i^m(t+1)$ for any $i\neq l_m^*$ as follows:
\begin{equation}
    \begin{array}{ll}
        (\alpha-\sum_{k\neq m}\sum_{j=l_k^*}^L z_j^k(t))p_m & \text{if} \ i=1  \\
        z_{i-1}^m(t) & \text{if} \ 1< i < l_m^* \\
        (1-p_m)z_{i-1}^m(t) & \text{if} \ l_m^*+1 < i <L\\ 
        (1-p_m) z_{L-1}^m(t)+(1-p_m)z_L^m(t) & \text{if} \ i=L \\
    \end{array}
\label{zmtplusone}
\end{equation}
Moreover, if $i=l_m^*+1$:
\begin{align}
z_{l_m^*+1}^m(t+1)=&-\sum_{j=1}^{l_m^*-1} z_j^m(t)-\sum_{j=l_m^*+1}^L (1-p_m) z_j^k(t)\nonumber\\&+\gamma_m-p_m(\alpha-\sum_{k\neq m}\sum_{j=l_k^*}^L z_j^k(t))
\label{zmlmbas}
\end{align} 
Based on this, we can conclude that $\widetilde{\boldsymbol{z}}(t+1)$ and $\widetilde{\boldsymbol{z}}(t)$ in the set $\jmath_{w^*}$ are related through the simpler linear equation:
\begin{equation}
\widetilde{\boldsymbol{z}}(t+1)=Q\widetilde{\boldsymbol{z}}(t)+\boldsymbol{c}
\label{linearrr}
\end{equation}
where $\widetilde{\boldsymbol{z}}(t)=[\widetilde{\boldsymbol{z}}^1(t),\ldots,\widetilde{\boldsymbol{z}}^K(t)]$, $\boldsymbol{c}\in\mathbb{R}^{(L-1)K}$ is a constant vector and $Q\in\mathbb{R}^{(L-1)K\times (L-1)K}$ is a square matrix that has the following form: \\       
\begin{equation}Q=\left[\begin{array}{ccccccc}
Q_1&0&\cdots &\cdots &\cdots &\cdots &0 \\
0&Q_2&\cdots &\cdots &\cdots &\cdots &0 \\
\vdots & &\ddots & & & & \\
A_1&A_2&\cdots &Q_m&\cdots&A_{K-1}&A_K\\
\vdots&  & &\ddots & &\vdots&\\
0&0&\cdots&\cdots&\cdots&Q_{K-1}&0\\
0&0&\cdots&\cdots&\cdots&0&Q_K\\
\end{array}\right]
\end{equation}
Based on eqs. (\ref{zktplusone})-(\ref{zmlmbas}), we can conclude the expressions of the matrices $Q_k$ for any $k\in\{1,\ldots,K\}$. These expressions are reported in Table \ref{tablematrices} of Appendix \ref{appendixmatrices}.
\subsection{Local optimality results}
Let us define $\boldsymbol{z}^*\in\mathcal{Z}$ as the system state vector that results from adopting the optimal policy of the relaxed problem. The exact expression of $\boldsymbol{z}^*$ can be concluded from (\ref{expressionofoptimalcost}). Specifically:
\begin{gather}
z^{*,k}_i=\gamma_k u_k^{l_k(W^*)}(i) \nonumber\\ z^{*,m}_i=\gamma_m[ \theta^*u_m^{l^2_m(W^*)}(i)+(1-\theta^*)u_m^{l^1_m(W^*)}(i)]
\end{gather}
Note that $\boldsymbol{z}^*$ belongs to the set $\jmath_{W^*}$. This can be seen by the definition of $\jmath_{W^*}$ as it coincides with the behavior of the optimal policy previously reported in Proposition \ref{optimalrelaxed}. Our goal in this section is to show that the system state vector $\boldsymbol{Z}^N(t)$ when the Whittle's index policy is adopted evolves closely to $\boldsymbol{z}^*$. In fact, when this happens, the achieved average age by the Whittle's index policy will be close to $C^{RP,N}$. To that end, we define the neighborhood set
\begin{equation}
\Omega_{\sigma}(\boldsymbol{z}^*)=\{\boldsymbol{z}\in\mathcal{Z}: ||\boldsymbol{z}-\boldsymbol{z}^*||\leq\sigma\}
\label{neighborhood}
\end{equation}
for any $\sigma>0$ where $||.||$ refers to the Euclidean distance. The next step of our analysis consists of laying out essential results on the matrix $Q$. 
\begin{definition}
Let $\lambda_1,\ldots,\lambda_N$ be the (real or complex) eigenvalues of a matrix $A\in\mathbb{C}^{N\times N}$, then its spectral radius $\rho(A)$ is defined as:
\begin{equation}
{\displaystyle \rho (A)=\max \left\{|\lambda _{1}|,\dotsc ,|\lambda _{n}|\right\}.}
\end{equation}
\end{definition}
\begin{theorem}
The spectral radius of the matrix $Q$, denoted by $\rho(Q)$, is strictly smaller than $1$.
\label{spectralradius}
\end{theorem}
\begin{IEEEproof}
The proof can be found in Appendix \ref{proofspectralradius}
\end{IEEEproof}
The above results will allow us to prove the convergence of the fluid limit $\boldsymbol{z}(t)$ to the optimal system state $\boldsymbol{z}^*$ of the RP as depicted in the next lemma.
\begin{lemma}
There exists $\sigma>0$ such that, if $\boldsymbol{z}(0)\in\Omega_{\sigma}(\boldsymbol{z}^*)\subseteq\jmath_{W^*}$, we have:
\begin{itemize}
\item $\boldsymbol{z}(t)\in \jmath_{W^*}\quad t\geq0$
\item $\boldsymbol{z}(t)\xrightarrow[t\to+\infty]{} \boldsymbol{z}^*$
\end{itemize}
\label{convergencezt}
\end{lemma}
\begin{IEEEproof}
The proof can be found in Appendix \ref{proofconvergencezt}.
\end{IEEEproof}
The above lemma tells us that the fluid limit vector $\boldsymbol{z}(t)$ converges to $\boldsymbol{z}^*$. In order to leverage these results to prove that $\boldsymbol{Z}^N(t)$ also converges to $\boldsymbol{z}^*$ when $N$ is large, we first characterize the gap between the system state $\boldsymbol{Z}^N(t)$ and the fluid limit $\boldsymbol{z}(t)$ in the following proposition. 
\begin{proposition}
There exists a neighborhood $\Omega_{\delta}(\boldsymbol{z}^*)$ such that, for any $T,\mu>0$, if $\boldsymbol{Z}^N(0)=\boldsymbol{x}\in \Omega_{\delta}(\boldsymbol{z}^*)$, there exists a constant $C_1$ independent of $N$ and $\boldsymbol{x}$ such that:
\begin{equation}
{\Pr}_{\boldsymbol{x}}(\underset{0\leq t<T}{\sup} ||\boldsymbol{Z}^N(t)-\boldsymbol{z}(t)||\geq\mu)\leq \frac{C_1}{N}
\end{equation}
where ${\Pr}_{\boldsymbol{x}}$ denotes the probability conditioned
on the initial state $\boldsymbol{Z}^N(0)=\boldsymbol{x}$.
\label{kurtzvariant}
\end{proposition}
\begin{IEEEproof}
The proof can be found in Appendix \ref{proofkurtzvariant}.
\end{IEEEproof}
By combining the results of Lemma \ref{convergencezt} and Proposition \ref{kurtzvariant}, we can now identify the gap between the system state vector $\boldsymbol{Z}^N(t)$ and $\boldsymbol{z}^*$. 
\begin{corollary}
There exists a neighborhood $\Omega_{\delta}(\boldsymbol{z}^*)$ such that, for any $\mu>0$, if $\boldsymbol{Z}^N(0)=\boldsymbol{x}\in \Omega_{\delta}(\boldsymbol{z}^*)$, then there exists a time $T_0$ such that for any time instant $T>T_0$, there exists a constant $C_{f}$ independent of $N$ and $\boldsymbol{x}$ such that:
\begin{equation}
{\Pr}_{\boldsymbol{x}}(\underset{T_0\leq t<T}{\sup} ||\boldsymbol{Z}^N(t)-\boldsymbol{z}^*||\geq\mu)\leq \frac{C_f}{N}
\end{equation} 
\label{corollaryforkurtz}
\end{corollary}
\begin{IEEEproof}
The proof can be found in Appendix \ref{proofcorollaryforkurtz}.
\end{IEEEproof}
Equipped with the above corollary, we can establish the local optimality of the Whittle's index policy. 
%
%
%
%
%
%
%
%
%
%
%
\begin{lemma}[Local Optimality]
Let $\{N_r\}_{r\in\mathbb{N}}$ be any increasing sequence of positive
integers such that $\alpha N_r, \gamma_kN_r \in \mathbb{N}$ for $k=1,\ldots,K$ and all $r$. When the Whittle's index policy is adopted, there exists a neighborhood $\Omega_{\delta}(\boldsymbol{z}^*)$ such that, if $\boldsymbol{Z}^{N_r}(0)=\boldsymbol{x}\in \Omega_{\delta}(\boldsymbol{z}^*)$, then
\begin{equation}
\lim_{T\to+\infty}\lim_{r\to+\infty}\frac{1}{T}\sum_{t=0}^{T-1}\mathbb{E}[v(\boldsymbol{Z}^{N_r}(t))]=C^{RP}
\end{equation}
where $v$ is a function that maps any system state $\boldsymbol{z}\in\mathcal{Z}$ to a per user average age value and $C^{RP}=\frac{C^{RP,N_r}}{N_r}$ is the optimal per user average age of the relaxed problem.
\label{asymptoticopt}
\end{lemma}
\begin{IEEEproof}
The proof can be found in Appendix \ref{proofasymptoticopt}.
\end{IEEEproof}
\begin{remark}
The sequence $\{N_r\}_{r\in\mathbb{N}}$ is used to ensure that the number of users within each class and the number of users that can transmit simultaneously are all integers.
\end{remark}
\section{Global optimality}
\begin{figure*}[ht]
\centering
\begin{subfigure}{0.33\textwidth}
  \centering
  \includegraphics[width=.99\linewidth]{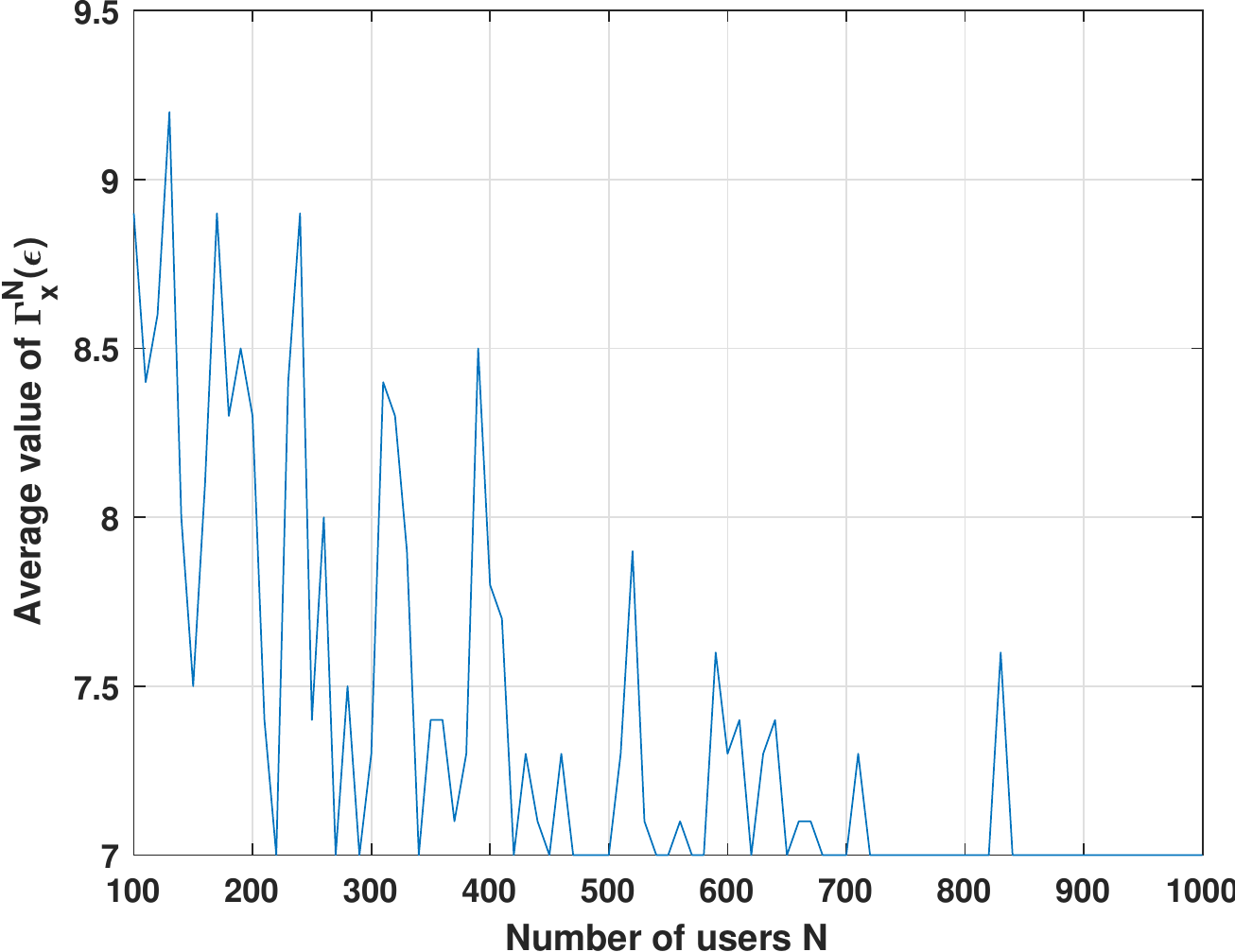}
  \caption{Average time $\Gamma^{N}_{\boldsymbol{x}}(\epsilon)$}
    \label{sim1}
\end{subfigure}%
\begin{subfigure}{0.33\textwidth}
\centering
  \includegraphics[width=.99\linewidth]{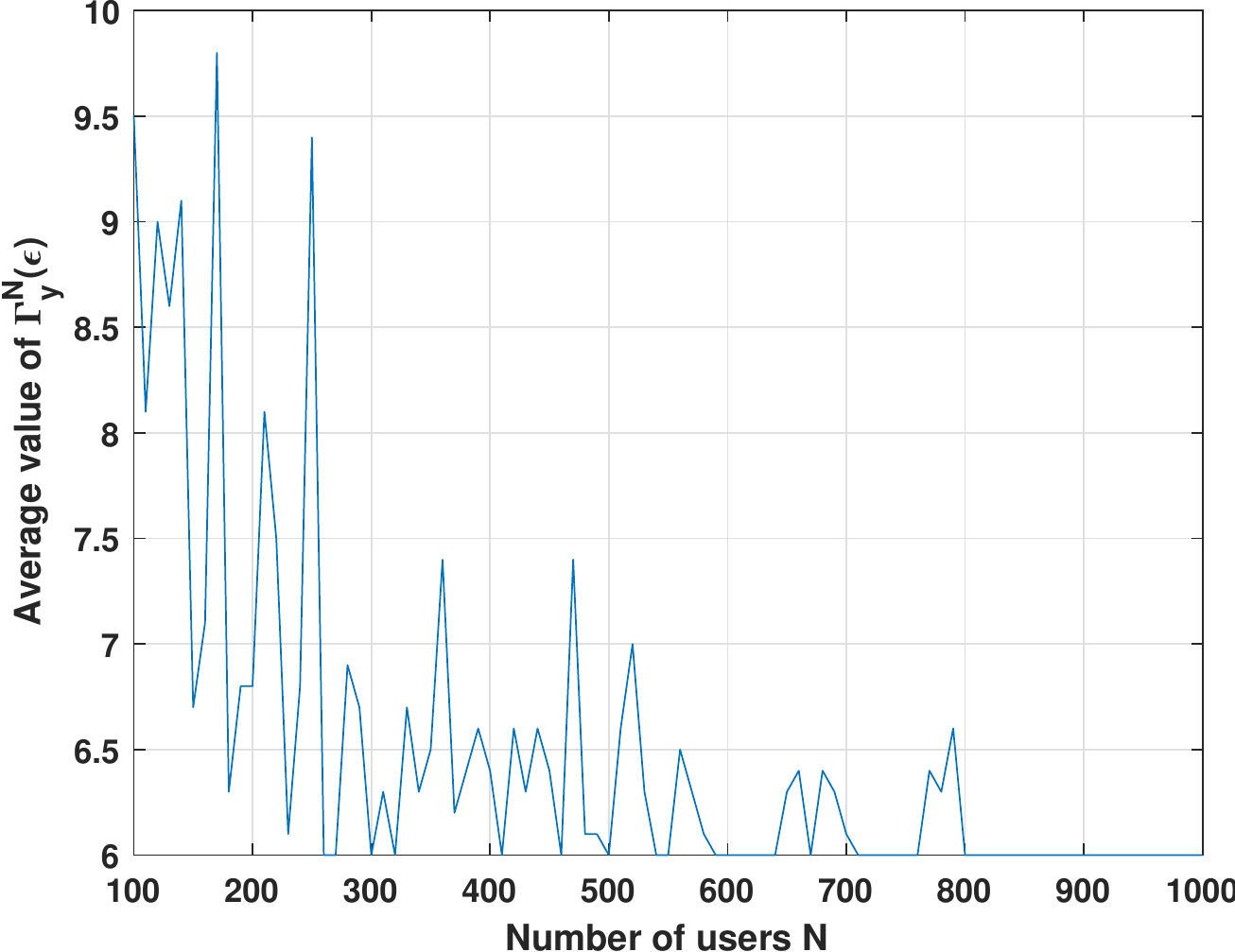}
  \caption{Average time $\Gamma^{N}_{\boldsymbol{y}}(\epsilon)$}
\label{sim2}
\end{subfigure}%
\begin{subfigure}{0.33\textwidth}
\centering
  \includegraphics[width=.99\linewidth]{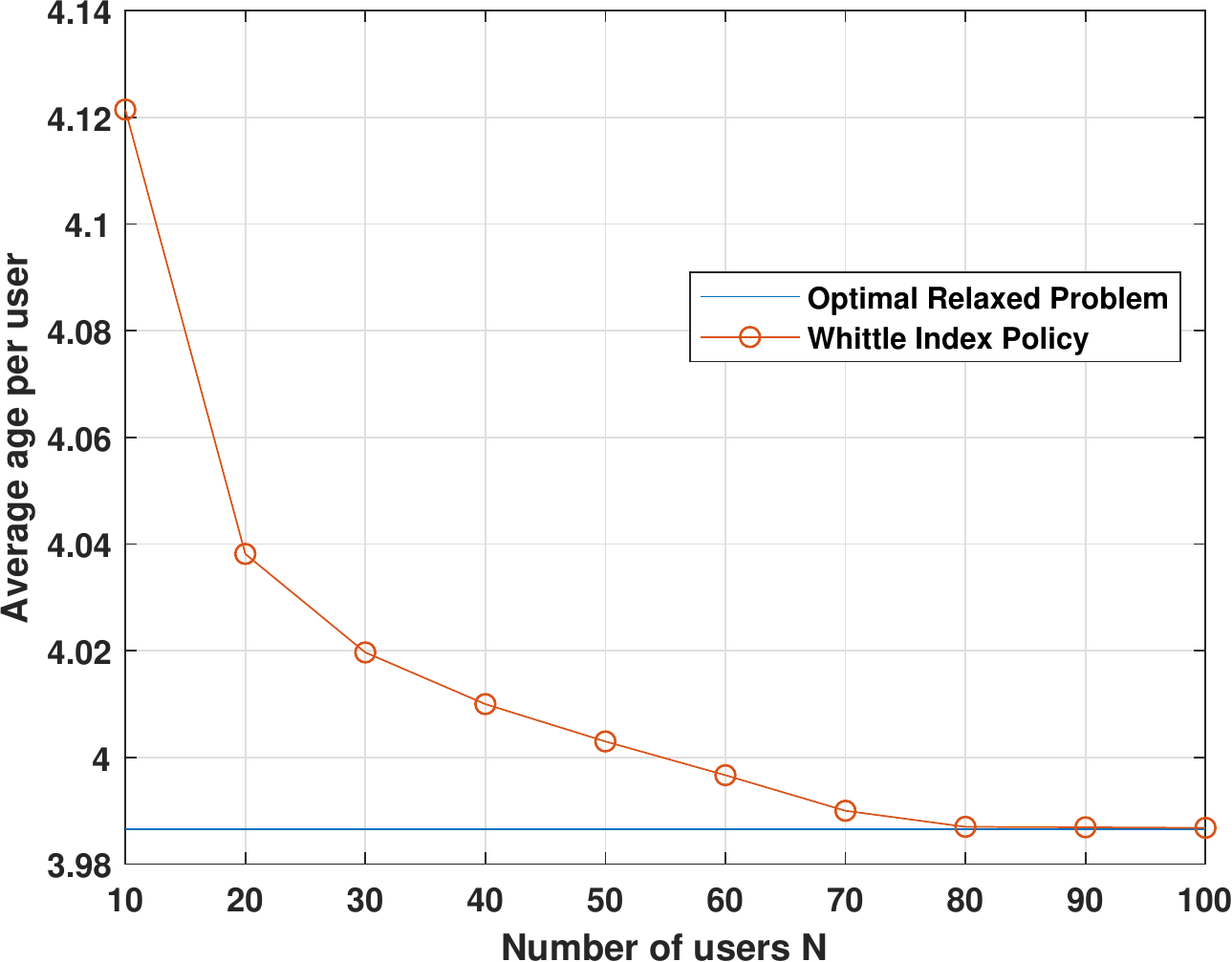}
  \caption{Average age per user comparison}
\label{sim3}
\end{subfigure}%
\caption{Simulation results}
\vspace{-20pt}
\end{figure*}
In this section, we extend our optimality results, under a recurrence assumption, to the case where $\boldsymbol{Z}^N(0)$ is an arbitrary initial system state. To proceed in this direction, we present the assumption that, coupled with our local asymptotic optimality results, will enable us to establish the global optimality of the Whittle's index scheduling policy. 
\begin{assumption}[\noindent\cite{2018arXiv180700352K,ouyang2016downlink,8115326}]
For any $\epsilon>0$, we let $\Omega_{\epsilon}(\boldsymbol{z}^*)$ be a neighborhood of the optimal system state $\boldsymbol{z}^*$. We suppose that this neighborhood is reachable from any initial state $\boldsymbol{Z}^N(0)=\boldsymbol{x}\in\mathcal{Z}$ and we define $\Gamma^{N}_{\boldsymbol{x}}(\epsilon)$ as the first time instant $t$ when the system state $\boldsymbol{Z}^N(t)\in\Omega_{\epsilon}(\boldsymbol{z}^*)$ starting from $\boldsymbol{Z}^N(0)=\boldsymbol{x}$. Specifically:
\begin{equation}
\Gamma^{N}_{\boldsymbol{x}}(\epsilon)=\min\{t:\boldsymbol{Z}^N(t)\in\Omega_{\epsilon}(\boldsymbol{z}^*)|\boldsymbol{Z}^N(0)=\boldsymbol{x}\}
\end{equation}
We assume that the expectation of $\Gamma^{N}_{\boldsymbol{x}}(\epsilon)$ is bounded:
\begin{equation}
\mathbb{E}[\Gamma^{N}_{\boldsymbol{x}}(\epsilon)]\leq M_{\epsilon}
\end{equation}
for all $\boldsymbol{x}$ and sufficiently large $N$.
\label{globaloptimalityassumption}
\end{assumption}
We can now proceed with establishing the global asymptotic optimality of the Whittle's index policy. For a number of users $N_r$, let us denote by $C_{\boldsymbol{x},\infty}^{WI,N_r}$ the infinite horizon cost of the Whittle's index policy starting from a system state $\boldsymbol{x}$. Specifically:
\begin{equation}
C_{\boldsymbol{x},\infty}^{WI,N_r}=\lim_{T\to+\infty}\frac{1}{T}\mathbb{E}\Big(\sum_{t=0}^{T-1}\sum_{k=1}^{K}\sum_{i=1}^{\gamma_kN_r}S_i^{k}(t)|\boldsymbol{Z}^{N_r}(0)=\boldsymbol{x}\Big)
\end{equation}
\begin{lemma}[Global Optimality]
Under Assumption \ref{globaloptimalityassumption}, and for any initial state $\boldsymbol{x}$, we have:
\begin{equation}
\lim_{r\to+\infty}\frac{C_{\boldsymbol{x},\infty}^{WI,N_r}}{N_r}=C^{RP}
\end{equation}
\label{ekhershiglobal}
\end{lemma}
\vspace{-15pt}
\begin{IEEEproof}
The proof can be found in Appendix \ref{proofekhershiglobal}.
\end{IEEEproof}

\section{Numerical Results}
\subsection{Verification of Assumption \protect\ref{globaloptimalityassumption} } 
In this section, we provide numerical results that showcase the validity of Assumption \ref{globaloptimalityassumption}. This assumption was considered in a variety of scheduling problems in the literature \cite{2018arXiv180700352K,ouyang2016downlink,8115326}.
%
In these references, Assumption \ref{globaloptimalityassumption} was shown to hold for different network parameters numerically. We proceed in the same direction in our work. In particular, we consider the following network settings:
\begin{itemize}
\item We consider two classes of users such that: $p_1=0.5$, $p_2=0.8$, $\gamma_1=0.5$ and $\gamma_2=0.5$.
\item The maximum age value $L$ is set to $200$.
\item The portion of users that can be scheduled at each time slot is $\alpha=0.5$.
\item The radius $\epsilon$ of the neighborhood $\Omega_{\epsilon}(\boldsymbol{z}^*)$ defined in (\ref{neighborhood}) is set to $0.04$.
\end{itemize}
We evaluate the average of the time $\Gamma^{N}_{\boldsymbol{z}_0}(\epsilon)$ after which $\boldsymbol{Z}^N(t)\in\Omega_{\epsilon}(\boldsymbol{z}^*)$ starting from the initial state $\boldsymbol{z}_0$. We select the two initial states to be two extreme points of the state space $\mathcal{Z}$ to showcase the uniformity of $\mathbb{E}[\Gamma^{N}_{\boldsymbol{z}_0}(\epsilon)]$ to the initial state. To that end, we let the initial states be: $\boldsymbol{x}=(\gamma_1,0,\ldots,0,\gamma_2,0,\ldots,0)$ and $\boldsymbol{y}=(0,\ldots,0,\gamma_1,0,\ldots,0,\gamma_2)$. In other words, all users initially have an age equal to $1$ and $L$ when $\boldsymbol{z}_0$ is equal to $\boldsymbol{x}$ and $\boldsymbol{y}$ respectively. We can see in Fig. \ref{sim1} and \ref{sim2} that even for a really small radius $0.04$, $\mathbb{E}[\Gamma^{N}_{\boldsymbol{z}_0}(\epsilon)]$ does not scale with $N$ and remains bounded for both initial states $\boldsymbol{x}$ and $\boldsymbol{y}$. Similar results can be found for different initial states and network settings, which put into perspective the validity of our assumption. 
\subsection{Implementation of the Whittle's index policy}
In this section, we implement the Whittle's index policy and evaluate its performance. To that end, we consider $N$ users in a cellular network where $\frac{N}{2}$ of them are cell-centered and, accordingly, have a probability of successful transmission $p_1=0.8$. The other $\frac{N}{2}$ are cell-edge users and, therefore, have a probability of successful transmission $p_1=0.2$.
%
%
%
%
We consider that the base station can schedule up to $M=\frac{N}{2}$ users at each time slot. We compare the per user average age of the Whittle's index policy to the optimal per user average age of the relaxed problem $C^{RP}$. As seen in Fig. \ref{sim3}, the performance gap between the two policies shrinks as the number of users increases, and vanishes for high values of $N$. These results corroborate our theoretical finding and showcases that the Whittle's index policy is indeed asymptotically optimal.  
%
%
%
%
\section{Conclusion}
In this paper, we have investigated the age performance of the Whittle's index policy in a network where $N$ users communicate with a central entity over unreliable channels. We have provided analytical results on its optimality in the many-users regime, a regime of particular interest in 5G networks. Numerical results were then presented that corroborate our theoretical findings and showcase the optimal performance of the policy when $N$ grows.
\bibliographystyle{IEEEtran}
\bibliography{trialout}
\appendices
\section{Proof of Theorem \protect\ref{increasingthreshold}}
\label{proofoftheoremincreasingthreshold}
The proof is based on studying the characteristics of the value function $V(.)$. Specifically, by proving that $V(.)$ is increasing, and by following an induction over time similar to the one done in \cite[Proposition~1]{2019arXiv190706604M},  we can show that the optimal policy is an increasing threshold policy. The details of this proof are omitted.
\section{Proof of Proposition \protect\ref{stationarydistribution}}
\label{proofofstationary}
It is sufficient to formulate the general balance equations at each state $i\in\{1,\ldots,L\}$, and knowing that $\sum\limits_{i=1}^{L} u^n(i)=1$, we can obtain the desired results. 
\section{Proof of Theorem \protect\ref{theoremcostgamma}}
\label{proofofcostgamma}
To obtain the results, we first point out that the cost incurred by being at state $S=i$ is nothing but the value $i$ of the state itself. Also, the user is scheduled whenever $S\geq n$. Accordingly, we have $\overline{C}(n,W)=\overline{C}_1(n)+\overline{C}_2(n,W)$ where: 
\begin{equation}
\overline{C}_1(n)=\lim_{T\to+\infty} \text{sup}\:\frac{1}{T}\sum_{t=0}^{T-1}S(t)=\sum\limits_{i=1}^{L}iu^n(i)
\end{equation}
\begin{equation}
\overline{C}_2(n,W)=\lim_{T\to+\infty} \text{sup}\:\frac{1}{T}\sum_{t=0}^{T-1}Wa(t)=W\sum\limits_{i=n}^{L}u^n(i) 
\end{equation}
By leveraging the results of Proposition \ref{stationarydistribution}, and after algebraic manipulations of the series involved, we can obtain the desired results.
\section{Proof of Proposition \protect\ref{indexabilityproperty}}
\label{proofofindexability}
We prove the indexability of the problem by leveraging the results of \cite[Proposition~2.2]{oatao14494}. Specifically, the proposition states that it is sufficient to prove that $\sum\limits_{i=n}^{L}u^n(i)$ is decreasing with $n$ for $p=p_1,\ldots,p_K$ to establish the indexability property for all classes. By using the expression of the stationary distribution provided in Proposition \ref{stationarydistribution}, we can show that $\sum\limits_{i=n}^{L}u^n(i)=\frac{1}{np+1-p}$ is decreasing with $n$ for any $p>0$. Since this is true for $p=p_1,\ldots,p_K$, we can conclude that the one-dimensional problem associated with any user of class $k$ is indexable.
\section{Proof of Proposition \protect\ref{closedformwhittle}}
\label{proofofclosedform}
As it was previously stated, the Whittle's index $W_i^k$ of a state $i$ of class $k$ is defined as the infimum subsidy $W$ that makes both idling and transmitting equally desirable in state $i$. Specifically:
\begin{equation}
W^k_i=\min\{W\in\mathbb{R}^+: i\in D^k(W)\}
\end{equation}
where $D^k(W)$ follows Definition \ref{indexabilitydefinition}. Using this definition to find the Whittle's index expressions can be challenging. To circumvent this, we first define the sequence $\big(W^k_i\big)_{i\in\{1,\ldots,L\}}$ as the intersection points between $\overline{C}(i,W)$ and $\overline{C}(i+1,W)$ when $p=p_k$. By leveraging the results in \cite[Corollary~2.1]{oatao14494}, we have that if $W^k_i$ is increasing with $i$, then the Whittle's index for any state $i$ of class $k$ is nothing but $W^k_i$. Consequently, we first seek a closed-form expression of $W^k_i$. By equating the two quantities $\overline{C}(i,W)$ and $\overline{C}(i+1,W)$, and by using the expressions reported in Theorem \ref{theoremcostgamma}, we get:
\begin{equation}
W^k_i=\frac{i(i-1)p_k}{2}+i-i(1-p_k)^{L-i}
\end{equation}
To pursue our analysis, we provide key results on the behavior of the intersection points.
\begin{proposition}
The sequence $\big(W^k_i\big)_{i\in\{1,\ldots,L\}}$ is increasing with $i$.
\label{propositionlambdaincrease}
\end{proposition}
\begin{IEEEproof}
To prove our claim, we consider the difference $\Delta(i)=W^k_{i+1}- W^k_i$. By using the expression of $W^k_i$, we can show that:
\begin{equation}
\Delta(i)=(ip_k+1)(1-(1-p_k)^{L-i-1})\geq0 \quad  i=1,\ldots,L-1
\end{equation}
for $k=1,\ldots,K$, which concludes our proof.
\end{IEEEproof}
Accordingly, we can conclude that the Whittle's index is $W^k_i$ for all states $i=1,\ldots,L$ and classes $k=1,\ldots,K$. 
\section{Proof of Proposition \protect\ref{lwvector}}
\label{proofoflwvector}
We follow in this proof the same line of work done in \cite[Appendix S]{2018arXiv180700352K}. To that end, for each class $k\in\{1,\ldots,K\}$, we distinguish between two cases:\\
\emph{1) There are no states $i\in\{1,\ldots,L\}$ such that $W_{i}^k=W$}: 
In this case, we point out that:
\begin{equation}
\underset{i}{\argmax} \{W_{i}^k:W_{i}^k\leq W\}=\underset{i}{\argmax} \{W_{i}^k:W_{i}^k<W\}
\end{equation}
In other words, the two expressions of $l_k(W)$ are equivalent. Now, let us consider a state $i\leq l_k(W)-1$. As it has been stated in Proposition \ref{propositionlambdaincrease}, $W_{i}^k$ is increasing in $i$. Accordingly, $W_{i}^k\leq W_{l_k(W)-1}^k<W$. Next, based on the indexability property of the problem, we can conclude that $D^k(W_{i}^k) \subseteq D^k(W)$. Note that by definition, $i\in D^k(W_{i}^k)$. Therefore, $i\in D^k(W)$. In other words, the optimal action for a user of class $k$ in state $i\leq l_k(W)-1$ is to stay idle. On the other hand, let us consider $q\geq l_{{k}}$. Knowing that there are no states $i$ such that $W_{i}^k\leq W$, and based on the fact that $l_k(W)-1=\underset{i}{\argmax} \{W_{i}^k:W_{i}^k<W\}$, we can conclude that $W^k_{l_k(W)}>W$. Therefore, $W<W_{q}^k$ with $W_{q}^k$ being equal to $\min\{\lambda\in\mathbb{R}^+: q\in D^k(\lambda)\}$ by definition. Accordingly, we can conclude that $q\not\in D^k(W)$ and the optimal action for a user of class $k$ in state $q\geq l_{k}(W)$ is to be scheduled for transmission. Consequently, $l_k(W)$ is indeed the optimal threshold. \\
\emph{2) There exists a state $f\in\{1,\ldots,L\}$ such that $W_{f}^k=W$}:
Let us consider any state $q\leq f$. By using Proposition \ref{propositionlambdaincrease}, we have $W_{q}^k\leq W_{f}^k=W$. Next, we leverage the indexability property of the problem to conclude that $D^k(W_{q}^k) \subseteq D^k(W)$. Accordingly, the optimal action when a user of class $k$ is in state $q$ is to stay idle. By considering any state $q\geq f+1>f$, and by employing the same argument, we can conclude that $W_{q}^k=\min\{\lambda\in\mathbb{R}^+: q\in D^k(\lambda)\}>W_{f}^k=W$. Accordingly, we can deduce that the optimal action for any user of class $k$ in state $q\geq f+1$ is to be scheduled for transmission. This showcases that $f+1$ is indeed an optimal threshold in this case. Note that $W_{f}^k$ was defined as the intersection between $\overline{C}(f,W)$ and $\overline{C}(f+1,W)$. Since $W_{f}^k=W$, we can conclude that both the thresholds $f$ and $f+1$ will lead to the same average cost in this case. Accordingly, $f$ and $f+1$ are both optimal thresholds. Since $f=\underset{i}{\argmax} \{W_{i}^k:W_{i}^k\leq W\}$, we can deduce that the optimal threshold $l_k(W)$ can indeed have one of the two expressions in (\ref{twoexpressions}).
\section{Proof of Proposition \protect\ref{optimalrelaxed}}
\label{proofoptimalrelaxed}
We recall that the Lagrangian approach can be summarized in the following problem:
\begin{equation}
 \underset{W\in\mathbb{R}^{+}}{\text{max}}\:\: \underset{\pi\in \Pi}{\text{min}}\:\:f(W,\pi)
\label{maxmin}
\end{equation}
From optimization theory, it is well-known that for any policy $\pi$, the problem in (\ref{maxmin}) forms a lower bound to our original problem in (\ref{relaxedobjective}). The difference between the two values is known as the \emph{duality} gap which is generally non-zero. We will prove in the following that we can achieve a zero duality gap in our case.

To proceed with our proof, we first recall that the optimal solution for a fixed $W$ of the problem (\ref{relaxedobjectivelagrange}) is a threshold policy with the threshold being $\boldsymbol{l}(W)$ provided in Proposition \ref{lwvector}. Suppose that there exists $W\in\mathbb{R}^+$ such that the constraint of (\ref{relaxedobjective}) is satisfied with equality. In other words:
\begin{equation}
\alpha=\sum_{k=1}^K \gamma_k\sum_{i=l_{k}(W^*)}^L u_k^{l_k(W^*)}(i)
\end{equation}
It is clear in this case that a threshold policy $\boldsymbol{l}(W^*)$ will be optimal for (\ref{relaxedobjective}) since:
\begin{equation}
\underset{W\in\mathbb{R}^{+}}{\text{max}}\:\: \underset{\pi\in \Pi}{\text{min}}\:\:f(W,\pi)=\overline{C}(\boldsymbol{l}(W^*),W^*)
\end{equation}
and the constraint in (\ref{relaxedobjective}) is satisfied. Accordingly, the duality gap will be zero. The issue is that $W^*$ does not necessarily exist. In fact, $\alpha$ is a real number that can take any value in $]0,1[$. On the other hand, the total proportion of scheduled users $A(W)$:
\begin{equation}
A(W)=\sum_{k=1}^K \gamma_k \sum_{i=l_{k}(W)}^L u^{l_k(W)}_k(i)
\end{equation}
is of discrete nature, since $\boldsymbol{l}(W)$ can only\footnote{We point out that threshold values higher than $L+1$ leads to the same average cost.} take values in $\{1,\ldots,L+1\}^K$. 

To address this issue, we leverage the results of Proposition \ref{lwvector} and define the following order relation in $\mathbb{R}^K$:
\begin{equation}
\forall W_1,W_2\in\mathbb{R}^+:\:\: W_1\leq W_2 , \quad\boldsymbol{l}(W_1)\leq \boldsymbol{l}(W_2)
\end{equation}
where $\boldsymbol{l}(W_1)\leq \boldsymbol{l}(W_2)$ if $l_k(W_1)\leq l_k(W_2)$ for $k=1,\ldots,K$. Therefore,  we can deduce that when $W$ is varied from the smallest Whittle's index value $W_1^{k}$ to the largest value $W_{L}^{k'}$, the optimal threshold vector varies from $l(W)=(1,\ldots,1)$ to $l(W)=(L+1,\ldots,L+1)$. We also point out that:
\begin{equation}
A(W)=\sum_{k=1}^K \gamma_k \sum_{i=l_k(W)}^L u^{l_k(W)}_k(i)=\sum_{k=1}^K \gamma_k \frac{W}{l_k(W)p+1-p}
\end{equation}
is decreasing with $l_k(W)$. Accordingly, $A(W)$ will vary from $1$ to $0$. Another result that the aforementioned proposition has highlighted is that when $W=W_{i}^{k}$ for any class $k$ and state $i$, there are two possible optimal threshold values that $l_k(W)$ can take. These two optimal thresholds $l^1_k(W)$ and $l^2_k(W)$ coincide when $W\neq W_i^k$ for any $i=1,\ldots,L$. On the other hand, these two thresholds verify the following inequality: $l^1_k(W)>l^2_k(W)$ when $W=W_i^k$ for a certain state $i$. By taking all this into account, we can deduce that there exists a class $m$ and state $p$ such that $A^2(W_p^m)\geq \alpha$ and $A^1(W_p^m)\leq \alpha$ by using the thresholds $l^2_m(W_p^m)$ and $l^1_m(W_p^m)$ for class $m$ respectively. If we let $W^*$ be equal to $W_p^m$, the optimal threshold for class $m$ can be either $l^1_m(W_p^m)$ or $l^2_m(W_p^m)$. We introduce a randomization parameter $\theta^*$ such that we use for class $m$ the threshold $l^2_m(W_p^m)$ with probability $\theta^*$ and $l^1_m(W_p^m)$ with probability $1-\theta^*$. By letting $\theta^*=\frac{\alpha-A^1(W_p^m)}{A^2(W_p^m)-A^1(W_p^m)}$, we can conclude that this randomized policy will satisfy the constraint in (\ref{relaxedobjective}) with equality. Accordingly, by using $W^*=W_p^m$ and the randomization factor $\theta^*$, the defined randomized policy will be optimal for the problem (\ref{relaxedobjective}). 

\section{Matrices expressions}
\label{appendixmatrices}
\begin{table*}[ht]
\centering
\begin{tabular}{p{0.99\linewidth}}
\begin{changemargin}{-2cm}{0 cm}
\[Q_k=
\begin{blockarray}{ccccccccccc}
 &1&2&\cdots&l_k^*-3&l_k^*-2&l_k^*&l_k^*+1&\cdots&L-1&L \\
\begin{block}{c(cccccccccc)}
  1&0& 0 &\cdots&0& 0 & p_k &p_k& \cdots &\cdots& p_k\\
  2 &1& \ddots& & & \vdots & 0 &0& &&0\\
  \vdots&0&\ddots&\ddots&&&\vdots&\vdots&&&\vdots&\\  
  l_k^*-3& 0 & \ddots& 1 & \ddots & \vdots &\vdots & \vdots & & &\vdots \\
  l_k^*-2& 0 & \cdots & 0 & 1 & 0 & 0& 0 &\cdots&\cdots&0\\
  l_k^* & -1 & \cdots &\cdots& \cdots &-1&-1&\cdots& \cdots &\cdots&-1 \\
  l_k^*+1 & 0 & 0 & \cdots &\cdots & 0 & 1-p_k&0&&0&0\\
  \vdots&\vdots&&&&\vdots&0&\ddots&\ddots&&0\\
  L-1&\vdots&&&&\vdots&&\ddots&\ddots&0&0\\
  L&0&\cdots&\cdots&\cdots&0&0&0&0&1-p_k&1-p_k\\  
\end{block}
\end{blockarray}
 \]
 \end{changemargin}
\begin{changemargin}{-2cm}{0 cm}
\[Q_m=
\begin{blockarray}{ccccccccccc}
 &1&2&\cdots&l_m-2&l_m-1&l_m+1&l_m+2&\cdots&L-1&L \\
\begin{block}{c(cccccccccc)}
  1&0& 0 &\cdots&0& 0 & 0 & 0 & \cdots &\cdots& 0\\
  2 &1& \ddots& & & \vdots & 0 &0& &&0\\
  \vdots&0&\ddots&\ddots&&&\vdots&\vdots&&&\vdots&\\  
  l_m-2& 0 & \ddots& 1 & \ddots & \vdots &\vdots & \vdots & & &\vdots \\
  l_m-1& 0 & \cdots & 0 & 1 & 0 & 0& 0 &\cdots&\cdots&0\\
  l_m+1 & -1 & \cdots &\cdots& \cdots &-1&p_m-1&\cdots& \cdots &\cdots&p_m-1 \\
  l_m+2 & 0 & 0 & \cdots &\cdots & 0 & 1-p_m&0&&0&0\\
  \vdots&\vdots&&&&\vdots&0&\ddots&\ddots&&0\\
  L-1&\vdots&&&&\vdots&&\ddots&\ddots&0&0\\
  L&0&\cdots&\cdots&\cdots&0&0&0&0&1-p_m&1-p_m\\  
\end{block}
\end{blockarray}
 \]
 \end{changemargin}
 \caption{The expressions of the matrices $Q_k$ for $k\neq m$ and $Q_m$}
 \label{tablematrices}
\end{tabular}
\label{allequations}
\hrulefill
\end{table*}
\section{Proof of Theorem \protect\ref{spectralradius}}
\label{proofspectralradius}
Our proof is based on finding the characteristic polynomial of $Q$, and investigating the norm of its roots. Contrary to existing works in this area in which the local optimality was simpler to establish, examining the matrix $Q$ is rather a difficult task in our case, as will be seen in the sequel. By examining the expression of $Q$, we can see that it can be written in the following block form:
\begin{equation}Q=\left[\begin{array}{cc}
\boldsymbol{F}_1&\boldsymbol{0}\\
\boldsymbol{0}&\boldsymbol{F}_2
\end{array}\right]
\end{equation}
where $\boldsymbol{F}_1\in\mathbb{R}^{(L-1)m\times (L-1)m}$ and $\boldsymbol{F}_2\in\mathbb{R}^{(L-1)(K-m)\times (L-1)(K-m)}$ are a lower triangular and upper triangular block matrices respectively. Accordingly, and by taking into account the expressions of $\boldsymbol{F}_1$ and $\boldsymbol{F}_2$, we can assert that the characteristic polynomial of $Q$ is the product of the characteristic polynomial of each matrix $Q_k$: 
\begin{equation}
\chi_{Q}(\lambda)=\prod_{k=1}^K \chi_{Q_k}(\lambda)
\end{equation}
Therefore, it is sufficient to find the eigenvalues of each matrix $Q_k$ to deduce those of $Q$. To that end, we distinguish between two cases:
\subsubsection{$k\neq m$} The characteristic polynomial of the matrix $Q_k$ is defined as follows:
\begin{equation}
\chi_{Q_k}=\det(Q_k-\lambda I)
\end{equation}
where $I\in\mathbb{R}^{(L-1)\times(L-1)}$ is the identity matrix. The characteristic polynomial of $Q_k$ is reported in Table \ref{tablepolynomials}. 
\begin{table*}[ht]
\centering
\begin{tabular}{p{0.99\linewidth}}
\begin{changemargin}{-2cm}{0 cm}
\[\chi_{Q_k}(\lambda)=
\begin{blockarray}{ccccccccccc}
 &1&2&\cdots&l_k^*-3&l_k^*-2&l_k^*&l_k^*+1&\cdots&L-1&L \\
\begin{block}{c|cccccccccc|}
  1&-\lambda& 0 &\cdots&0& 0 & p_k &p_k& \cdots &\cdots& p_k\\
  2 &1& -\lambda& & & \vdots & 0 &0& &&0\\
  \vdots&0&\ddots&\ddots&&&\vdots&\vdots&&&\vdots&\\  
  l_k^*-3& 0 & \ddots& 1 & \ddots & \vdots &\vdots & \vdots & & &\vdots \\
  l_k^*-2& 0 & \cdots & 0 & 1 & -\lambda & 0& 0 &\cdots&\cdots&0\\
  l_k^* & -1 & \cdots &\cdots& \cdots &-1&-1-\lambda&\cdots& \cdots &\cdots&-1 \\
  l_k^*+1 & 0 & 0 & \cdots &\cdots & 0 & 1-p_k&-\lambda&&0&0\\
  \vdots&\vdots&&&&\vdots&0&\ddots&\ddots&&0\\
  L-1&\vdots&&&&\vdots&&\ddots&\ddots&-\lambda&0\\
  L&0&\cdots&\cdots&\cdots&0&0&0&0&1-p_k&1-p_k-\lambda\\  
\end{block}
\end{blockarray}
 \]
 \end{changemargin}
\[\boldsymbol{G}=
\left(\begin{array}{@{}c|c@{}}
  \begin{matrix}
  -\lambda& 0 &\cdots&0& 0 \\
  0& -\lambda& & & \vdots\\
  0&\ddots&\ddots&&\vdots\\  
  0 & \ddots& 0 & \ddots & \vdots \\
  0 & \cdots & 0 & 0 & -\lambda \\
  \end{matrix}
  & \bigb \\
\hline
  \bigzero &
  \begin{matrix}
  -\lambda-1-\frac{1}{\lambda} \sum_{i=0}^{l_k^*-2} \frac{p_k}{\lambda^i}& -1-\frac{1}{\lambda} \sum_{i=0}^{l_k^*-2} \frac{p_k}{\lambda^i}&\cdots& \cdots& \cdots& -1-\frac{1}{\lambda} \sum_{i=0}^{l_k^*-2} \frac{p_k}{\lambda^i}\\
  1-p_k& -\lambda&0 &\cdots&\cdots& 0\\
  0 & \ddots &\ddots& & &\vdots\\
  \vdots&&\ddots&\ddots& &\vdots\\
  \vdots&&&&-\lambda&0\\
  0&\cdots&\cdots&\cdots&1-p_k&1-p_k-\lambda
  \end{matrix}
\end{array}\right)
\] 
 \caption{The expressions of the characteristic polynomials of $Q_k$}
 \label{tablepolynomials}
\end{tabular}
\label{whatever1}
\hrulefill
\end{table*}
In order to get a closed-form of this determinant, we apply elementary row and column operations. More specifically, let us denote by $r_i$ the row $i$ of the determinant. We also denote by $a_{i,j}$ the element in row $i$ and column $j$ of the matrix $Q_k$. For $i=2$ till $l_k^*-2$, we add to each row the previous row multiplied by $\frac{1}{\lambda}$. In other words: 
\begin{equation}
r_i=\frac{r_{i-1}}{\lambda}+r_i \quad i=2,\ldots,l_k^*-2
\end{equation}
Note that these operations are not done simultaneously but rather successively. In other words, in the increasing order of the rows, and at each iteration $i$, we add to $r_i$ the updated row $r_{i-1}$ at iteration $i-1$ multiplied by $\frac{1}{\lambda}$. After doing so, we execute the following operation in order to have zeros for the elements $a_{l_k^*,1}$ to $a_{l_k^*,l_k^*-2}$: 
\begin{equation}
r_{l_k^*}=-\sum_{i=1}^{l_k^*-2}\frac{r_i}{\lambda} + r_{l_k^*}
\end{equation}  
As a result, $\chi_{Q_k}(\lambda)$ will be the determinant of the matrix $\boldsymbol{G}$ reported in the same table. Since $\boldsymbol{G}$ is an upper triangular block matrix, we will not be interested in the expression of $\boldsymbol{B}$ as the determinant will be independent of it. By letting $A=1+\frac{1}{\lambda} \sum_{i=0}^{l_k^*-3} \frac{p_k}{\lambda^i}$, the determinant of $\boldsymbol{G}$ will be equal to $(-\lambda)^{l_k^*-2}$ (the determinant of the upper left block) times the following determinant:
\[  S_k=
\begin{blockarray}{|cccccc|}
  -\lambda-A& -A&\cdots& \cdots& \cdots& -A\\
  1-p_k& -\lambda&0 &\cdots&\cdots& 0\\
  0 & \ddots &\ddots& & &\vdots\\
  \vdots&&\ddots&\ddots& &\vdots\\
  \vdots&&&&-\lambda&0\\
  0&\cdots&\cdots&\cdots&1-p_k&1-p_k-\lambda
 \end{blockarray}
\]
which originates from the lower right block matrix of $\boldsymbol{G}$, and of dimension $(L-l_k^*+1)\times(L-l_k^*+1)$. Now, we need to find an explicit expression of this determinant, which we will denote by $S_k$. To achieve this goal, we first start by developing the determinant through the last column. By doing so, we end up with:
\begin{equation} 
S_k=(-1)^{L-l_k^*} (-A) (1-p_k)^{L-l_k^*}+(-\lambda+1-p_k)D_{L-l_k^*}
\end{equation}
where $D_n$ is determinant of the following matrix:
\[  
\begin{blockarray}{cccccc}
   &1&2&\cdots&\cdots&n\\
  \begin{block}{c(ccccc)}
  1&-\lambda-A& -A&\cdots& \cdots& -A\\
  2&1-p_k& -\lambda&0 &\cdots& 0\\
  \vdots& 0 & \ddots &\ddots& &\vdots\\
  \vdots&&\ddots&\ddots& &0\\
  n&0& \cdots&\cdots&1-p_k&-\lambda\\
  \end{block}
 \end{blockarray}
\]  
We need to find the expression of $D_n$ for any fixed integer value $n$ in order to conclude it for $D_{L-l_k^*}$. To do so, we define $\Delta_n$ as: 
\[ \Delta_n= 
\begin{blockarray}{|ccccc|}
  -\lambda-A& -A&\cdots& \cdots& -A\\
  1& -\frac{\lambda}{1-p_k}&0 &\cdots& 0\\
  0 & \ddots &\ddots& &\vdots\\
  &\ddots&\ddots& &0\\
  0& \cdots&\cdots&1&-\frac{\lambda}{1-p_k}\\
 \end{blockarray}
\]   
Hence, $D_n=(1-p_k)^{n-1}\Delta_n$.\\
The second step consists of computing $\Delta_n$ for all $n$. To that end, we provide the following lemma.
\begin{lemma}
The determinant $\Delta_n$ can be expressed for any integer value $n$ as:
\begin{equation}
\Delta_n=(-1)^n \sum_{j=0}^{n-2} (\frac{\lambda}{1-p_k})^j A+(-1)^{n+1} (\frac{\lambda}{1-p_k})^{n-1} \Delta_1 \quad n\geq2
\label{expression_delta}
\end{equation} 
where $\Delta_1=-\lambda-A$.
\end{lemma}
\begin{IEEEproof}
The proof follows a mathematical induction. For $n=1$, we can easily check that $\Delta_1$ is indeed $-\lambda-A$ by replacing $n$ by $1$. Next, we suppose that (\ref{expression_delta}) holds up till $n$ and we aim to prove that it holds for $n+1$. By developing $\Delta_{n+1}$ through the last column, we get:
\begin{equation}
\Delta_{n+1}=(-1)^{n+2}(-A)-\frac{\lambda}{1-p_k}\Delta_n
\end{equation}
By replacing $\Delta_n$ with its value, we end up with: 
\begin{align*}
&\Delta_{n+1}=(-1)^{n+2}(-A)-\frac{\lambda}{1-p_k}[(-1)^n \sum_{j=0}^{n-2} (\frac{\lambda}{1-p_k})^j A\\ & +(-1)^{n+1} (\frac{\lambda}{1-p_k})^{n-1} \Delta_1]=(-1)^{n+2}(-A)+(-1)^{n+1}\\ & \sum_{j=0}^{n-2} (\frac{\lambda}{1-p_k})^{j+1} A+(-1)^{n} (\frac{\lambda}{1-p_k})^{n} \Delta_1=(-1)^{n+2}(-A)\\
&+(-1)^{n+1} \sum_{j=1}^{n-1} (\frac{\lambda}{1-p_k})^{j} A+(-1)^{n} (\frac{\lambda}{1-p_k})^{n} \Delta_1\\
&=(-1)^{n+1}A+(-1)^{n+1} \sum_{j=1}^{n-1} (\frac{\lambda}{1-p_k})^{j} A+(-1)^{n} (\frac{\lambda}{1-p_k})^{n} \Delta_1\\
&=(-1)^{n+1} \sum_{j=0}^{n-1} (\frac{\lambda}{1-p_k})^{j} A+(-1)^{n+2} (\frac{\lambda}{1-p_k})^{n} \Delta_1
\end{align*} 
Therefore, (\ref{expression_delta}) holds for $n+1$ which concludes our proof. 
\end{IEEEproof}
By leveraging the above lemma, we can conclude the expression of $D_{L-l_k^*}$:
\begin{align}
D_{L-l_k^*}=&(1-p_k)^{L-l_k^*-1}[(-1)^{L-l_k^*} \sum_{j=0}^{L-l_k^*-2} (\frac{\lambda}{1-p_k})^j A\nonumber\\&+(-1)^{L-l_k^*+1} (\frac{\lambda}{1-p_k})^{L-l_k^*-1} \Delta_1]
\end{align}
As a consequence, we are able to find $S_k$ in function of $\lambda$ and $p_k$, as well as the expression of $\chi_{Q_k}(\lambda)$.
In fact, after computations, we get: 
\begin{equation}
\chi_{Q_k}(\lambda)=(-1)^{L+1} \lambda^{L-l_k^*} (\lambda^{l_k^*-1}+\sum_{i=0}^{l_k^*-2} p_k \lambda^i)
\end{equation}
Now, based on the expression of the characteristic polynomial $\chi_{Q_k}$, we prove that all the eigenvalues of $Q_k$ have a modulus strictly less than one. We prove this result by contradiction. More specifically, we suppose there exists a given eigenvalue of the matrix $Q_k$ that satisfies $|\lambda|\geq 1$. As $\lambda$ is an eigenvalue of $Q_k$, it is therefore a root of $\chi_{Q_k}(\lambda)$. Hence, it verifies:
\begin{equation}
\lambda^{l_k^*-1}=-\sum_{i=0}^{l_k^*-2} p_k \lambda^i=-p_k \frac{1-\lambda^{l_k^*-1}}{1-\lambda}
\end{equation}
By factorizing the element $\lambda^{l_k^*}$, and by using the modulus on both sides, we get:
\begin{align*}
p_k&=|\lambda|^{l_k^*-1}|p_k-1+\lambda|\\
&\geq^{(a)} |\lambda|^{l_k^*-1}(|\lambda|-|1-p_k|)\\
& \geq^{(b)} |\lambda|(|\lambda|-|1-p_k|)\\
& = |\lambda|^2- |\lambda|(1-p_k)
\end{align*}
where $(a)$ and $(b)$ originate from the reverse triangular inequality and the fact that $|\lambda|\geq 1$ respectively. Hence:
\begin{equation}
|\lambda|^2- |\lambda|(1-p_k)-p_k \leq 0
\end{equation}
By employing standard real functions analysis, it can be shown that the polynomial:
\begin{equation}
x^2-(1-p_k)x-p_k
\end{equation}
is negative if and only if $x\in [-p_k,1]$. However, $|\lambda| \geq 1$ by assumption. Accordingly, $|\lambda|$ can only be equal to $1$. Next, we prove that, in this case, the imaginary part of $\lambda$ is equal to zero. To that end, let us consider $\lambda=x+iy$. Therefore, we have:
\begin{equation}
p_k=|\lambda|^{l_k^*-1} |p_k-1+x+iy|=|p_k-1+x+iy|
\end{equation}
By using the definition of the modulus, and by squaring both sides, we get:
\begin{equation}
p_k^2=(1-x-p_k)^2+y^2
\end{equation}
Knowing that $x^2+y^2=1$, we can deduce: 
\begin{equation}
\frac{2-2p_k}{2(1-p_k)}=x
\end{equation}  
Hence, $x=1$, i.e. $y=0$, and we can deduce that $\lambda=1$. 
However, $1$ is not eigenvalue of matrix $Q_k$. This can be seen by replacing $\lambda$ with $1$ in the characteristic polynomial of $Q_k$. Accordingly, the hypothesis that $|\lambda| \geq 1$ fails and all the eigenvalues of $Q_k$ for any $k \neq m$ have a modulus strictly than $1$.
\subsubsection{$k=m$} 
The characteristic polynomial of the matrix $Q_m$ is reported in Table \ref{tableform}.
\begin{table*}[ht]
\centering
\begin{tabular}{p{0.99\linewidth}}
\begin{changemargin}{-2cm}{0 cm}
\[\chi_{Q_m}(\lambda)=
\begin{blockarray}{ccccccccccc}
 &1&2&\cdots&l_m^*-2&l_m^*-1&l_m^*+1&l_m^*+2&\cdots&L-1&L \\
\begin{block}{c|cccccccccc|}
  1&-\lambda& 0 &\cdots&0& 0 & 0 &0& \cdots &\cdots& 0\\
  2 &1& -\lambda& & & \vdots & 0 &0& &&0\\
  \vdots&0&\ddots&\ddots&&&\vdots&\vdots&&&\vdots&\\  
  l_m^*-2& 0 & \ddots& 1 & \ddots & \vdots &\vdots & \vdots & & &\vdots \\
  l_m^*-1& 0 & \cdots & 0 & 1 & -\lambda & 0& 0 &\cdots&\cdots&0\\
  l_m^*+1 & -1 & \cdots &\cdots& \cdots &-1&p_m-1-\lambda&\cdots& \cdots &\cdots&p_m-1 \\
  l_m^*+2 & 0 & 0 & \cdots &\cdots & 0 & 1-p_m&-\lambda&&0&0\\
  \vdots&\vdots&&&&\vdots&0&\ddots&\ddots&&0\\
  L-1&\vdots&&&&\vdots&&\ddots&\ddots&-\lambda&0\\
  L&0&\cdots&\cdots&\cdots&0&0&0&0&1-p_m&1-p_m-\lambda\\  
\end{block}
\end{blockarray}
 \]
 \end{changemargin}
 \newcommand{\bigc}{\mbox{\normalfont\Large\bfseries C}}
\[\boldsymbol{H}=
\left(\begin{array}{@{}c|c@{}}
  \begin{matrix}
  -\lambda& 0 &\cdots&0& 0 \\
  0& -\lambda& & & \vdots\\
  0&\ddots&\ddots&&\vdots\\  
  0 & \ddots& 0 & \ddots & \vdots \\
  0 & \cdots & 0 & 0 & -\lambda \\
  \end{matrix}
  & \bigc \\
\hline
  \bigzero &
  \begin{matrix}
  -\lambda-1+p_m& -1+p_m&\cdots& \cdots& \cdots& -1+p_m\\
  1-p_m& -\lambda&0 &\cdots&\cdots& 0\\
  0 & \ddots &\ddots& & &\vdots\\
  \vdots&&\ddots&\ddots& &\vdots\\
  \vdots&&&&-\lambda&0\\
  0&\cdots&\cdots&\cdots&1-p_m&1-p_m-\lambda
  \end{matrix}
\end{array}\right)
\]
 \caption{The expressions of the characteristic polynomials of $Q_m$}
 \label{tableform}
\end{tabular}
\label{whatever2}
\hrulefill
\end{table*}
We follow the same steps of the previous case. For $i=2$ till $l_m^*-1$, we sequentially add to each row, the previous row multiplied by $\frac{1}{\lambda}$. In other words: 
\begin{equation}
r_i=\frac{r_{i-1}}{\lambda}+r_i \quad i=2,\ldots,l_m^*-1
\end{equation}
Then, we execute the following operation in order to have zeros for the elements $a_{l_m^*+1,1}$ to $a_{l_m^*+1,l_m^*-1}$: 
\begin{equation}
r_{l_m^*+1}=-\sum_{i=1}^{l_m^*-1} \frac{r_i}{\lambda} + r_{l_m^*+1}
\end{equation}
As a result, $\chi_{Q_m}(\lambda)$ will be the determinant of the matrix $\boldsymbol{H}$ reported in the same table. By replacing $1-p_m$ with $A$, the determinant of $\boldsymbol{H}$ will be equal to $(-\lambda)^{l_m^*-1}$ multiplied by the following determinant: 
\[  
\begin{blockarray}{|cccccc|}
  -\lambda-A& -A&\cdots& \cdots& \cdots& -A\\
  1-p_m& -\lambda&0 &\cdots&\cdots& 0\\
  0 & \ddots &\ddots& & &\vdots\\
  \vdots&&\ddots&\ddots& &\vdots\\
  \vdots&&&&-\lambda&0\\
  0&\cdots&\cdots&\cdots&1-p_m&1-p_m-\lambda
 \end{blockarray}
\]   
Therefore, we end up with the same determinant $S_k$ defined previously but with $A=1-p_m$. Fortunately, we have already computed this determinant for any $k$ and regardless of the value of $A$. Given that, we can find the expression of $S_m$ in function of $\lambda$ and $p_m$ as well as the expression of $\chi_{Q_m}(\lambda)$. After extensive computations, we obtain:
\begin{equation}
\chi_{Q_m}(\lambda)=(-\lambda)^{L-1}
\end{equation} 
Hence, the only eigenvalue of $Q_m$ is $0$ that has a multiplicity of $L-1$.

By combining all these results, we can conclude that the eigenvalues of the matrix $Q$ have a modulus strictly less than one. Accordingly, the spectral radius of the matrix $Q$, denoted by $\rho(Q)$, is strictly less than $1$ which concludes our proof.
\section{Proof of Lemma \protect\ref{convergencezt}}
\label{proofconvergencezt}
The first part of our proof consists of identifying the fixed points of the fluid limit model reported in (\ref{zelases}). To that end, we provide the following lemma.
\begin{lemma}
The optimal state vector of the RP $\boldsymbol{z}^*$ is the unique fixed point of the fluid limit model. In other words, for all $\boldsymbol{z}\in\mathcal{Z}$, $\boldsymbol{z}=Q'(\boldsymbol{z})\boldsymbol{z}$ if and only if $\boldsymbol{z}=\boldsymbol{z}^*$.
\end{lemma}
\begin{IEEEproof}
The proof follows the same methodology of \cite{weber1990index} and \cite[Lemma~9]{ouyang2016downlink}.
\end{IEEEproof}
With the above lemma in mind, what remains is to show the convergence of the fluid model to the fixed point $\boldsymbol{z}^*$. By leveraging the results of Section IV-C, we can assert that there exists a neighborhood of $\boldsymbol{z}^*$, denoted by $\Omega_{\sigma_0}(\boldsymbol{z}^*)\subseteq\jmath_{W^*}$, such that the fluid limit model is linear and follows (\ref{linearrr}). To that end, let us define $\widetilde{\boldsymbol{z}}^*=[\widetilde{\boldsymbol{z}}^{*,1}(t),\ldots,\widetilde{\boldsymbol{z}}^{*,K}(t)]$ in a similar fashion to what we have previously done for $\widetilde{\boldsymbol{z}}(t)$ in Section IV-C. Accordingly, in this neighborhood
\begin{equation}
\widetilde{\boldsymbol{z}}(t)-\widetilde{\boldsymbol{z}}^*=Q^t(\widetilde{\boldsymbol{z}}(0)-\widetilde{\boldsymbol{z}}^*)
\end{equation} 
We recall that the spectral radius of the matrix $Q$, denoted by $\rho(Q)$, was shown  to be strictly smaller than $1$. Consequently, according to the stability theory of linear
systems, $\widetilde{\boldsymbol{z}}(t)$ will converge to $\widetilde{\boldsymbol{z}}^*$ if the initial state is close enough to $\widetilde{\boldsymbol{z}}^*$. Therefore, there exists $\sigma<\sigma_0$ such that
if $\boldsymbol{z}(0)\in\Omega_{\sigma}(\boldsymbol{z}^*)\subseteq\jmath_{W^*}$, $\boldsymbol{z}(t)\in \jmath_{W^*}$ for any $t\geq0$, and
$\boldsymbol{z}(t)\xrightarrow[t\to+\infty]{} \boldsymbol{z}^*$.
\section{Proof of Proposition \protect\ref{kurtzvariant}}
\label{proofkurtzvariant}
To prove this proposition, we first consider a general time instant $t\in\{0,\ldots,T\}$ and provide the following lemma.
\begin{lemma}
There exists a neighborhood $\Omega_{\epsilon}(\boldsymbol{z}^*)$ such that, for any $\mu>0$, if $\boldsymbol{Z}^N(t)=\boldsymbol{z}\in\Omega_{\epsilon}(\boldsymbol{z}^*)$, there exists a constant $C$ independent of $N$ and $\boldsymbol{z}$ such that:
\begin{equation}
\Pr(||\boldsymbol{Z}^N(t+1)-Q'(\boldsymbol{z})\boldsymbol{z}|\boldsymbol{Z}^N(t)=\boldsymbol{z}||\geq\mu|\boldsymbol{Z}^N(t)=\boldsymbol{z})\leq \frac{C}{N}
\end{equation}
\label{onetimeinstant}
\end{lemma}
\vspace{-15pt}
\begin{IEEEproof}
To proceed with our proof, we first recall that:
\begin{equation}
\boldsymbol{Z}^N(t)=(\boldsymbol{Z}^{1,N}(t),\ldots,\boldsymbol{Z}^{K,N}(t))
\end{equation}
where $\boldsymbol{Z}^{k,N}(t)=(Z_1^{k,N}(t),\ldots,Z_L^{k,N}(t))$ is the state vector for a specific class $k$. With that in mind, we point out that $\{\boldsymbol{Z}^N(t+1): ||\boldsymbol{Z}^N(t+1)-Q'(\boldsymbol{z})\boldsymbol{z}||\geq\mu|\boldsymbol{Z}^N(t)=\boldsymbol{z}\}\subseteq \bigcup\limits_{i,k}^{}\{\boldsymbol{Z}^N(t+1): ||Z^{k,N}_i(t+1)-[Q'(\boldsymbol{z})\boldsymbol{z}]^k_i||\geq\frac{\mu}{KL}|\boldsymbol{Z}^N(t)=\boldsymbol{z}\}$ where $[Q'(\boldsymbol{z})\boldsymbol{z}]^k_i$ is the class $k$ and state $i$ component of $Q'(\boldsymbol{z})\boldsymbol{z}$. Accordingly, the following holds: $\Pr(||\boldsymbol{Z}^N(t+1)-Q'(\boldsymbol{z})\boldsymbol{z}||\geq\mu|\boldsymbol{Z}^N(t)=\boldsymbol{z})\leq \sum_{i,k}^{}\Pr(|Z^{k,N}_i(t+1)-[Q'(\boldsymbol{z})\boldsymbol{z}]^k_i|\geq \frac{\mu}{KL}|\boldsymbol{Z}^N(t)=\boldsymbol{z})$.

\noindent Next, we recall that, by definition:
\begin{equation}
Q'(\boldsymbol{z})\boldsymbol{z}=\mathbb{E}[\boldsymbol{Z}^N(t+1)|\boldsymbol{Z}^N(t)=\boldsymbol{z}]
\label{expectationequality}
\end{equation}
The above equality will allow us to leverage the Chebyshev's inequality to find the desired results. 
%
%
%
Let us consider $\boldsymbol{Z}^N(t)=\boldsymbol{z}\in\Omega_{\epsilon}(\boldsymbol{z}^*)\subseteq\jmath_{W^*}$. The evolution of $\boldsymbol{Z}^N(t)$ in this neighborhood was previously defined in Section IV-C. Specifically, users with a Whittle's index larger than $W^*$ are scheduled while users with an index strictly smaller than $W^*$ are left to idle. Users with an index equal to $W^*$ are scheduled with a certain randomization parameter. To that end, let us consider a constant $\mu>0$ and a class $k$ among the available classes. By considering the threshold $l_k^*$ depicted in Proposition \ref{lwvector}, we have the following:
\begin{itemize}
\item For any state $2\leq i\leq l_k^*$:
\begin{equation}
Z^{k,N}_i(t+1)=Z^{k,N}_{i-1}(t)
\end{equation}
Accordingly, by using (\ref{expectationequality}), we have: $[Q'(\boldsymbol{z})\boldsymbol{z}]^k_i=Z^{k,N}_i(t+1)=Z^{k,N}_{i-1}(t)$. Therefore, $\Pr(||Z^{k,N}_i(t+1)-[Q'(\boldsymbol{z})\boldsymbol{z}]^k_i||\geq \frac{\mu}{KL}|\boldsymbol{Z}^N(t)=\boldsymbol{z})=0$.
\item For state $i=1$, we denote by $\alpha_k$ the scheduled portion of users belonging to class $k$. Accordingly, we have:
\begin{equation}
NZ^{k,N}_1(t+1)|\boldsymbol{Z}^N(t)=\boldsymbol{z}\sim {\rm Binomial}(N\alpha_k,p_k)
\end{equation}
Consequently:
\begin{equation}
\mathrm{Var}[NZ^{k,N}_1(t+1)|\boldsymbol{Z}^N(t)=\boldsymbol{z}]=p_k(1-p_k)\alpha_k N
\end{equation}
By dividing over $N$, and by leveraging the Chebyshev's inequality, we get:
\begin{align}
&\Pr(|Z^{k,N}_1(t+1)-[Q'(\boldsymbol{z})\boldsymbol{z}]^k_1|\geq \frac{\mu}{KL}|\boldsymbol{Z}^N(t)=\boldsymbol{z})\nonumber\\&\leq  \frac{p_k(1-p_k)\alpha_k K^2L^2}{N\mu^2}
\end{align}
\item For any state $l_k^*+1< i<L$, we have:
\begin{equation}
NZ^{k,N}_i(t+1)|\boldsymbol{Z}^N(t)=\boldsymbol{z}\sim {\rm Binomial}(NZ^{k,N}_{i-1}(t),1-p_k)
\end{equation}
As we have previously done, we can conclude:
\begin{align}
&\Pr(|Z^{k,N}_i(t+1)-[Q'(\boldsymbol{z})\boldsymbol{z}]^k_i|\geq \frac{\mu}{KL}|\boldsymbol{Z}^N(t)=\boldsymbol{z})\nonumber\\&\leq  \frac{p_k(1-p_k)Z^{k,N}_{i-1}(t) K^2L^2}{N\mu^2}
\end{align}
\item $i=L:$ by proceeding in a similar way to the previous case, we can deduce:
\begin{align}
&\Pr(|Z^{k,N}_L(t+1)-[Q'(\boldsymbol{z})\boldsymbol{z}]^k_L|\geq \frac{\mu}{KL}|\boldsymbol{Z}^N(t)=\boldsymbol{z})\nonumber\\&\leq  \frac{p_k(1-p_k)(Z^{k,N}_{L-1}(t)+Z^{k,N}_{L}(t)) K^2L^2}{N\mu^2}
\end{align}
\item For $i=l_k^*+1$, we distinguish between two scenarios: 1) $k\neq m:$ in this case, the evolution is similar to the case where $i>l_k^*+1$, 2) $k=m:$ in this case, the portion $Z^{m,N}_{l^2_m}(t)$ is scheduled with a probability $\theta$. Therefore, by proceeding in the same way as before, we get:
\begin{align}
&\Pr(|Z^{m,N}_{l^2_m+1}(t+1)-[Q'(\boldsymbol{z})\boldsymbol{z}]^m_{l^2_m+1}|\geq \frac{\mu}{KL}|\boldsymbol{Z}^N(t)=\boldsymbol{z})\nonumber\\&\leq  \frac{p_m(1-p_m)(\theta Z^{m,N}_{l^2_m}(t))K^2L^2}{N\mu^2}
\end{align}
\end{itemize}
By combining all the above results, and by leveraging the union inequality previously provided along with the fact that $\alpha_k\leq1$, $\theta\leq1$, and $\sum_{i,k}^{}Z^{k,N}_{i}(t)\leq 1$, we can conclude:
\begin{align}
&\Pr(||\boldsymbol{Z}^N(t+1)-Q'(\boldsymbol{z})\boldsymbol{z}|\boldsymbol{Z}^N(t)=\boldsymbol{z}||\geq\mu|\boldsymbol{Z}^N(t)=\boldsymbol{z})\nonumber\\&\leq 
\sum_{k=1}^K \frac{p_k(1-p_k)K^2L^2}{N\mu^2}=\frac{C}{N}
\end{align}
where $C=\frac{\sum_{k=1}^Kp_k(1-p_k)K^2L^2}{\mu^2}$ which concludes our proof.
\end{IEEEproof}
The next step of our proof is to use the above results to evaluate the difference between the fluid limit and the system state vector for any time instant $t\geq1$. To do so, we provide the following lemma.
\begin{lemma}
There exists a neighborhood $\Omega_{\delta}(\boldsymbol{z}^*)$ such that, for any $\mu>0$, if $\boldsymbol{Z}^N(0)=\boldsymbol{x}\in\Omega_{\delta}(\boldsymbol{z}^*)$, then for any $t\geq1$, there exists a constant $c_1^t$ independent of $N$ and $\boldsymbol{x}$ such that:
\begin{equation}
{\Pr}_{\boldsymbol{x}}(||\boldsymbol{Z}^N(t)-\boldsymbol{z}(t)||\geq\mu)\leq \frac{c_1^t}{N}
\end{equation}
\end{lemma}
\begin{IEEEproof}
The proof of this lemma can be obtained by slightly adapting the results obtained in \cite[Lemma~18]{ouyang2016downlink}. Let us consider $\nu<\mu$. We recall from Lemma \ref{onetimeinstant} that there exists a neighborhood $\Omega_{\epsilon}(\boldsymbol{z}^*)$ such that, for any $\mu>0$, if $\boldsymbol{z}\in\Omega_{\epsilon}(\boldsymbol{z}^*)$, there exists a constant $C$ independent of $N$ and $\boldsymbol{z}$ such that:
\begin{equation}
\Pr(||\boldsymbol{Z}^N(t+1)-Q'(\boldsymbol{z})\boldsymbol{z}|\boldsymbol{Z}^N(t)=\boldsymbol{z}||\geq\mu|\boldsymbol{Z}^N(t)=\boldsymbol{z})\leq \frac{C}{N}
\label{recalllemma}
\end{equation}
We let $\rho<\epsilon$ be such that:
\begin{equation}
||Q'(\boldsymbol{x})\boldsymbol{x}-Q'(\boldsymbol{y})\boldsymbol{y}||\leq\nu
\label{inequalityimportant}
\end{equation}
for all $\boldsymbol{x},\boldsymbol{y}$ with $||\boldsymbol{x}-\boldsymbol{y}||\leq\rho$. This is possible since the function: $ \boldsymbol{z} \to Q'(\boldsymbol{z})\boldsymbol{z}$ is linear, and accordingly, Lipschitz continuous. Next, we recall the definition of $\sigma$ in
Lemma \ref{convergencezt}. To that end, we let $\delta<\min(\sigma,\epsilon)$ be such that, if $\boldsymbol{z}(0)\in\Omega_{\delta}(\boldsymbol{z}^*)$, then $\boldsymbol{z}(t)\in\Omega_{\epsilon-\rho}(\boldsymbol{z}^*)$ for $t\geq1$. With the above parameters specified, we prove the statement by a mathematical induction. 

For $t=1$, and since $\boldsymbol{x}\in\Omega_{\delta}(\boldsymbol{z}^*)\subseteq\Omega_{\epsilon}(\boldsymbol{z}^*)$, the following holds:
\begin{align}
{\Pr}_{\boldsymbol{x}}(||\boldsymbol{Z}^N(1)-\boldsymbol{z}(1)||\geq\mu)
=&{\Pr}_{\boldsymbol{x}}(||\boldsymbol{Z}^N(1)-Q'(\boldsymbol{x})\boldsymbol{x}||\geq\mu)\nonumber\\&\leq\frac{C}{N}
\end{align}
and the desired results hold for $t=1$ by simply choosing $c^1_1=C$. Let us suppose that the statement holds for any $t\geq1$. We investigate the property for $t+1$. To that end:
\begin{align}
&{\Pr}_{\boldsymbol{x}}(||\boldsymbol{Z}^N(t+1)-\boldsymbol{z}(t+1)||\geq\mu)\nonumber\\&={\Pr}_{\boldsymbol{x}}(||\boldsymbol{Z}^N(t+1)-\boldsymbol{z}(t+1)||\geq\mu\Big|||\boldsymbol{Z}^N(t)-\boldsymbol{z}(t)||\geq\rho)\nonumber\\&{\Pr}_{\boldsymbol{x}}(||\boldsymbol{Z}^N(t)-\boldsymbol{z}(t)||\geq\rho)+\nonumber\\&{\Pr}_{\boldsymbol{x}}(||\boldsymbol{Z}^N(t+1)-\boldsymbol{z}(t+1)||\geq\mu\Big|||\boldsymbol{Z}^N(t)-\boldsymbol{z}(t)||<\rho)\nonumber\\&{\Pr}_{\boldsymbol{x}}(||\boldsymbol{Z}^N(t)-\boldsymbol{z}(t)||<\rho)\leq^{(a)} \frac{d^t_1}{N}+\nonumber\\&{\Pr}_{\boldsymbol{x}}(||\boldsymbol{Z}^N(t+1)-\boldsymbol{z}(t+1)||\geq\mu\Big|||\boldsymbol{Z}^N(t)-\boldsymbol{z}(t)||<\rho)
\label{firststepinit}
\end{align}
where $(a)$ follows from ${\Pr}_{\boldsymbol{x}}(||\boldsymbol{Z}^N(t+1)-\boldsymbol{z}(t+1)||\geq\mu\Big|||\boldsymbol{Z}^N(t)-\boldsymbol{z}(t)||\geq\rho)\leq1$ and $d^t_1$ being the constant related to the statement holding for $t$ for $\rho$. Next, we tackle the second term of the inequality in (\ref{firststepinit}):
\begin{align}
&{\Pr}_{\boldsymbol{x}}(||\boldsymbol{Z}^N(t+1)-\boldsymbol{z}(t+1)||\geq\mu\Big|||\boldsymbol{Z}^N(t)-\boldsymbol{z}(t)||<\rho)\nonumber\\&={\Pr}_{\boldsymbol{x}}(||\boldsymbol{Z}^N(t+1)-Q'(\boldsymbol{Z}^N(t))\boldsymbol{Z}^N(t)+Q'(\boldsymbol{Z}^N(t))\boldsymbol{Z}^N(t)\nonumber\\&-\boldsymbol{z}(t+1)||\geq\mu\Big|||\boldsymbol{Z}^N(t)-\boldsymbol{z}(t)||<\rho)\nonumber\\
&\leq^{(a)}{\Pr}_{\boldsymbol{x}}(||\boldsymbol{Z}^N(t+1)-Q'(\boldsymbol{Z}^N(t))\boldsymbol{Z}^N(t)||+||Q'(\boldsymbol{Z}^N(t))\nonumber\\&\boldsymbol{Z}^N(t)-Q'(\boldsymbol{z}(t))\boldsymbol{z}(t)||\geq\mu\Big|||\boldsymbol{Z}^N(t)-\boldsymbol{z}(t)||<\rho)\nonumber\\
&\leq^{(b)}{\Pr}_{\boldsymbol{x}}(||\boldsymbol{Z}^N(t+1)-Q'(\boldsymbol{Z}^N(t))\boldsymbol{Z}^N(t)||\geq\mu-\nu\nonumber\\& \Big|||\boldsymbol{Z}^N(t)-\boldsymbol{z}(t)||<\rho)\nonumber\\&=\sum_{\boldsymbol{z}\in\Omega_{\rho}(\boldsymbol{z}(t))}{\Pr}_{\boldsymbol{x}}(\boldsymbol{Z}^N(t)=\boldsymbol{z}\Big|\boldsymbol{Z}^N(t)\in\Omega_{\rho}(\boldsymbol{z}(t)))\nonumber\\&{\Pr}_{\boldsymbol{x}}(||\boldsymbol{Z}^N(t+1)-Q'(\boldsymbol{z})\boldsymbol{z}||\geq\mu-\nu|\boldsymbol{Z}^N(t)=\boldsymbol{z})
\label{secondstepinit}
\end{align}
where $(a)$ and $(b)$ follows from the triangular inequality and the relationship in (\ref{inequalityimportant}). Since $z(t)\in\Omega_{\epsilon-\mu}(\boldsymbol{z}^*))$ and $\rho<\epsilon$, we have $\Omega_{\rho}(\boldsymbol{z}(t))\subseteq\Omega_{\epsilon}(\boldsymbol{z}^*))$. Accordingly, from eq. (\ref{recalllemma}), we can conclude:
\begin{align}
&{\Pr}_{\boldsymbol{x}}(||\boldsymbol{Z}^N(t+1)-Q'(\boldsymbol{z})\boldsymbol{z}||\geq\mu-\nu|\boldsymbol{Z}^N(t)=\boldsymbol{z})\leq \frac{C'}{N}
\end{align}
where $C'=\frac{\sum_{k=1}^Kp_k(1-p_k)K^2L^2}{(\mu-\nu)^2}$. By substituting the above results in (\ref{secondstepinit}), we get:
\begin{multline}
{\Pr}_{\boldsymbol{x}}(||\boldsymbol{Z}^N(t+1)-\boldsymbol{z}(t+1)||\geq\mu\Big|||\boldsymbol{Z}^N(t)-\boldsymbol{z}(t)||<\rho)\leq \frac{C'}{N}
\end{multline}
Combining this with (\ref{firststepinit}), we can conclude that there exists a constant $c_1^{t+1}$ such that:
\begin{equation}
{\Pr}_{\boldsymbol{x}}(||\boldsymbol{Z}^N(t+1)-\boldsymbol{z}(t+1)||\geq\mu)\leq \frac{c_1^{t+1}}{N}
\end{equation}
which concludes our inductive proof.
\end{IEEEproof}
We recall that from the union bound, 
\begin{align}
&\Pr(\underset{0\leq t<T}{\sup} ||\boldsymbol{Z}^N(t)-\boldsymbol{z}(t)||\geq\mu)\nonumber\\&\leq\sum_{t=0}^{T-1}{\Pr}_{\boldsymbol{x}}(||\boldsymbol{Z}^N(t)-\boldsymbol{z}(t)||\geq\mu)
\end{align}
Accordingly, from the previous lemma, and over finite time horizon
$T$, there exists a constants $C_1$ that is independent of $\boldsymbol{x}$ and $N$ such that:
\begin{equation}
\Pr(\underset{0\leq t<T}{\sup} ||\boldsymbol{Z}^N(t)-\boldsymbol{z}(t)||\geq\mu)\leq \frac{C_1}{N}
\end{equation}  
which concludes our proof.
\section{Proof of Corollary \protect\ref{corollaryforkurtz}}
\label{proofcorollaryforkurtz}
We let $0<\nu<\mu$. We recall from Proposition \ref{kurtzvariant} that $\delta<\sigma$ and, accordingly, from Lemma \ref{convergencezt}, given $\boldsymbol{Z}^N(0)=\boldsymbol{x}\in \Omega_{\delta}(\boldsymbol{z}^*)$, there exists $T_0$ such that for any $t\geq T_0$:
\begin{equation}
||\boldsymbol{z}(t)-\boldsymbol{z}^*||\leq\nu
\end{equation}
By leveraging Proposition \ref{kurtzvariant}, we have:
\begin{align}
&{\Pr}_{\boldsymbol{x}}(\underset{T_0\leq t<T}{\sup} ||\boldsymbol{Z}^N(t)-\boldsymbol{z}^*||\geq\mu)
\nonumber\\&
\leq {\Pr}_{\boldsymbol{x}}(\underset{T_0\leq t<T}{\sup} ||\boldsymbol{Z}^N(t)-\boldsymbol{z}(t)||+||\boldsymbol{z}(t)-\boldsymbol{z}^*||\geq\mu)\nonumber\\&\leq {\Pr}_{\boldsymbol{x}}(\underset{T_0\leq t<T}{\sup} ||\boldsymbol{Z}^N(t)-\boldsymbol{z}(t)||\geq\mu-\nu)\nonumber\\&
\leq
{\Pr}_{\boldsymbol{x}}(\underset{0\leq t<T}{\sup} ||\boldsymbol{Z}^N(t)-\boldsymbol{z}(t)||\geq\mu-\nu)\leq \frac{C_f}{N}
\end{align}
which concludes the proof.
\section{Proof of Lemma \protect\ref{asymptoticopt}}
\label{proofasymptoticopt}
The proof of this lemma can be obtained by slightly adapting the results obtained in \cite[p.~18]{ouyang2016downlink}. To prove our lemma, let us first denote by $v$ the function that maps any system state $\boldsymbol{z}\in\mathcal{Z}$ to a per user average age value. Specifically, $v: \boldsymbol{z} \to\sum_{k=1}^{K}\sum_{i=1}^{L}iz^k_i$. Accordingly, $Nv(\boldsymbol{Z}^N(t))$ is the instantaneous total age of users in the network at time $t$ when the Whittle's index policy is employed and the number of users is $N$. Moreover, $v(\boldsymbol{z}^*)=C^{RP}=\frac{C^{RP,N}}{N}$ is the optimal per user average age of the relaxed problem. For each $l>0$, we let $\mu>0$ be such that for any $\boldsymbol{x}\in\mathcal{Z}$, if $||\boldsymbol{x}-\boldsymbol{z}^*||\leq\mu$, then:
\begin{equation}
|v(\boldsymbol{x})-C^{RP}|\leq l
\label{inequalityossetl}
\end{equation}
This is possible since the function $v(.)$ is linear and, accordingly, Lipschitz continuous. Let us consider a time horizon $T$, a number of users $N_r$, and the time instant $T_0$ of Corollary \ref{corollaryforkurtz}. To establish the desired optimality, we evaluate the following difference:
\begin{align}
&|\frac{1}{T}\sum_{t=0}^{T_0-1}\mathbb{E}[v(\boldsymbol{Z}^{N_r}(t)))-C^{RP}]+\nonumber\\&\frac{1}{T}\sum_{t=T_0}^{T-1}\mathbb{E}[v(\boldsymbol{Z}^{N_r}(t))-C^{RP}]|\nonumber\\
&\leq^{(a)}|\frac{1}{T}\sum_{t=0}^{T_0-1}\mathbb{E}[v(\boldsymbol{Z}^{N_r}(t)))-C^{RP}]|+ \nonumber\\&|\frac{1}{T}\sum_{t=T_0}^{T-1}\mathbb{E}[v(\boldsymbol{Z}^{N_r}(t))-C^{RP}]|\leq^{(b)} \frac{LT_0}{T} + \nonumber\\&\frac{1}{T}\sum_{t=T_0}^{T-1}\mathbb{E}[|v(\boldsymbol{Z}^{N_r}(t))-C^{RP}|]
\end{align}
where $(a)$ follows from the triangular inequality. On the other hand, since $v((\boldsymbol{Z}^{N_r}(t))\leq L$ and $C^{RP}\geq0$, $(b)$ follows from the fact that $v((\boldsymbol{Z}^{N_r}(t))-C^{RP}\leq L$. We now tackle the second term of the inequality. To that end, we let $A_r$ be the event $\{\sup_{T_0\leq t<T} ||\boldsymbol{Z}^{N_r}(t)-\boldsymbol{z}^*||\geq\mu)\}$. We can rewrite the second term as:
\begin{align}
&
{\Pr}_{\boldsymbol{x}}(A_r)\frac{1}{T}\sum_{t=T_0}^{T-1}\mathbb{E}[|v(\boldsymbol{Z}^{N_r}(t))-C^{RP}|\big|A_r]+\nonumber\\&(1-{\Pr}_{\boldsymbol{x}}(A_r))\frac{1}{T}\sum_{t=T_0}^{T-1}\mathbb{E}[|v(\boldsymbol{Z}^{N_r}(t))-C^{RP}|\big|\overline{A}_r]\nonumber\\ & \leq^{(a)}\frac{L(T-T_0)}{T}{\Pr}_{\boldsymbol{x}}(A_r)+(1-{\Pr}_{\boldsymbol{x}}(A_r))l
\end{align}
where $(a)$ follows from (\ref{inequalityossetl}) and that $v((\boldsymbol{Z}^{N_r}(t))-C^{RP}\leq L$. By leveraging Corollary \ref{corollaryforkurtz}, we know if $\boldsymbol{Z}^{N_r}(0)=\boldsymbol{x}\in \Omega_{\delta}(\boldsymbol{z}^*)$, ${\Pr}_{\boldsymbol{x}}(A_r)\xrightarrow[r\to+\infty]{}0$. Accordingly, 
\begin{equation}
\lim_{r\to+\infty} |\frac{1}{T}\sum_{t=0}^{T-1}\mathbb{E}[v(\boldsymbol{Z}^{N_r}(t)))-C^{RP}]|\leq \frac{LT_0}{T}+l
\end{equation}
As $l$ is an arbitrarily positive value, and by letting $T\rightarrow+\infty$, we can conclude that:
\begin{equation}
\lim_{T\to+\infty}\lim_{r\to+\infty}\frac{1}{T}\sum_{t=0}^{T-1}\mathbb{E}[v(\boldsymbol{Z}^{N_r}(t))]=C^{RP}
\end{equation}
given $\boldsymbol{Z}^{N_r}(0)=\boldsymbol{x}\in \Omega_{\delta}(\boldsymbol{z}^*)$.
\vspace{-5pt}
\section{Proof of Lemma \protect\ref{ekhershiglobal}}
\label{proofekhershiglobal}
We first provide in the following lemma a characterization of the behavior of $\boldsymbol{Z}^N(t)$.
\begin{lemma}
Under the Whittle's index policy, the system state vector $\boldsymbol{Z}^N(t)$ evolves as an aperiodic Markov chain with only one recurrent class.
\label{recurrencechain}
\end{lemma}
\begin{IEEEproof}
To prove this lemma, it is sufficient to establish that there exists a state $\boldsymbol{z}_f$ that is reachable from any initial state $\boldsymbol{Z}^N(0)=\boldsymbol{x}\in\mathcal{Z}$. Let us consider the following possible event $B$ which can happen with a probability $\Pr(B)>0$: for $L$ consecutive time slots, every scheduled user suffers from a failed transmission. When the event $B$ takes place, the system state, regardless of its initial value, will become equal to $\boldsymbol{z}_f=(\boldsymbol{z}_f^1,\ldots,\boldsymbol{z}_f^K)$ where $\boldsymbol{z}_f^k=(0,\ldots,1)$. In other words, all users will have an age that is equal to $L$. This implies that state $\boldsymbol{z}_f$ is reachable from any other state which concludes the proof.
\end{IEEEproof}
Equipped with the above lemma, and by noting that the system's state space is finite, we can conclude that there exists a steady-state distribution for the system state vector $\boldsymbol{Z}^N(t)$ under the Whittle's index policy. We denote this distribution by $\boldsymbol{Z}^N(\infty)$. Next, we lay out the following lemma.
\begin{lemma}
Under Assumption \ref{globaloptimalityassumption}, the steady-state distribution vector verifies:
\begin{equation}
\lim_{r\to+\infty}\Pr(\boldsymbol{Z}^{N_r}(\infty)\in\Omega_{\epsilon}(\boldsymbol{z}^*))=1
\end{equation}
\label{abelekhershi}
\end{lemma}
\vspace{-15pt}
\begin{IEEEproof}
The proof follows the same argument provided in \cite[Lemma~6]{ouyang2016downlink}. Specifically, the idea of the proof revolves around investigating the random variables $T_{\epsilon}$ and $T_{\epsilon}^{0}$ defined as the time between consecutive hitting times into the neighborhood $\Omega_{\epsilon}(\boldsymbol{z}^*)$ and the sojourn time in $\Omega_{\epsilon}(\boldsymbol{z}^*)$ respectively. By leveraging Assumption \ref{globaloptimalityassumption} and Corollary \ref{corollaryforkurtz}, we can show that, as $N$ grows, the expected portion of time spent outside the neighborhood $\Omega_{\epsilon}(\boldsymbol{z}^*)$ vanishes.
\end{IEEEproof}
Equipped with the above lemma, we can proceed with proving our desired results. Similarly, for each $l>0$, we let $\epsilon>0$ be such that for any $\boldsymbol{z}\in\mathcal{Z}$, if $||\boldsymbol{x}-\boldsymbol{z}^*||\leq\epsilon$, then:
\begin{equation}
|v(\boldsymbol{z})-v(\boldsymbol{z}^*)|\leq l
\label{inequalityossetl2}
\end{equation}
We recall that $C^{RP}=v(\boldsymbol{z}^*)$. We define the event $E_r$ as $\{\boldsymbol{Z}^{N_r}(\infty)\in\Omega_{\epsilon}(\boldsymbol{z}^*)\}$. Next, we investigate the following difference:
\begin{align}
&|\frac{C_{\boldsymbol{x},\infty}^{WI,N_r}}{N_r}-C^{RP}|\leq^{(a)} \mathbb{E}[|v(\boldsymbol{Z}^{N_r}(\infty))-C^{RP}|]\nonumber\\&
=\Pr(E_r)\mathbb{E}[|v(\boldsymbol{Z}^{N_r}(\infty))-C^{RP}|\big|E_r]+\nonumber\\&\Pr(\overline{E}_r)\mathbb{E}[|v(\boldsymbol{Z}^{N_r}(\infty))-C^{RP}|\big|\overline{E}_r]\leq l\Pr(E_r)+\Pr(\overline{E}_r)L
\end{align}
where $(a)$ follows from Jensen's inequality. Using the results of Lemma \ref{abelekhershi}, and knowing that $l$ is arbitrarily small, we can conclude:
\begin{equation}
\lim_{r\to+\infty}\frac{C_{\boldsymbol{x},\infty}^{WI,N_r}}{N_r}=C^{RP}
\end{equation}
\end{document}